\documentclass[10pt, conference, compsocconf]{IEEEtran}
\usepackage{amsmath,amssymb,amsfonts}
\usepackage[hyphens]{url}

\def\BibTeX{{\rm B\kern-.05em{\sc i\kern-.025em b}\kern-.08em
    T\kern-.1667em\lower.7ex\hbox{E}\kern-.125emX}}

\newlength{\bibitemsep}
\newlength{\bibparskip}
\let\oldthebibliography\thebibliography
\renewcommand\thebibliography[1]{%
  \oldthebibliography{#1}%
  \setlength{\parskip}{\bibitemsep}%
  \setlength{\itemsep}{\bibparskip}%
}

\usepackage{xspace}
\newcommand*{\Scale}[2][4]{\scalebox{#1}{$#2$}}
\usepackage{hhline}

\let\labelindent\relax

\usepackage{enumitem}

\usepackage{makecell}
\usepackage[utf8]{inputenc}
\usepackage{xfrac}
\usepackage{balance}

\usepackage{multirow}
\usepackage{multicol}
\usepackage{tabu}
\usepackage[T1]{fontenc}
\usepackage{mathtools}

\newcolumntype{P}[1]{>{\centering\arraybackslash}p{#1}}
\newcolumntype{L}[1]{>{\raggedright\let\newline\\\arraybackslash\hspace{0pt}}m{#1}}
\newcolumntype{C}[1]{>{\centering\let\newline\\\arraybackslash\hspace{0pt}}m{#1}}
\newcolumntype{R}[1]{>{\raggedleft\let\newline\\\arraybackslash\hspace{0pt}}m{#1}}

\newcommand*{\Comb}[2]{{}_{#1}C_{#2}}%

\newcommand{\maxcount}{M\xspace}

\newcommand{\name}{Mithril\xspace}
\newcommand{\nameplus}{Mithril+\xspace}
\newcommand{\pararfm}{PARFM\xspace}
\newcommand{\rh}{Row Hammer\xspace}

\newcommand{\maxptr}{MaxPtr\xspace}
\newcommand{\minptr}{MinPtr\xspace}

\newcommand{\nth}{Flip_{TH}\xspace}
\newcommand{\adth}{Ad_{TH}\xspace}
\newcommand{\rfmth}{RFM_{TH}\xspace}

\newcommand{\entrynum}{N_{entry}\xspace}

\usepackage{dblfloatfix}

\title{Mithril: Cooperative Row Hammer Protection on Commodity DRAM Leveraging Managed Refresh}

\author{
Michael Jaemin Kim$^{\dag}$\quad~Jaehyun Park$^{\dag}$\quad~Yeonhong Park$^{\ddag}$\quad~Wanju Doh$^{\S}$\quad~Namhoon Kim$^{\dag}$\\
Tae Jun Ham$^{\ddag}$\quad~Jae W. Lee$^{\ddag}$\quad~Jung Ho Ahn$^{\dag\S}$\\
\begin{tabular}{c}
Dept. of \{Intelligence and Information$^{\dag}$, Computer Science and Engineering$^{\ddag}$\}, Prog. in Artificial Intelligence$^{\S}$\\
Seoul National University\\
\{michael604, wogus20002, ilil96, wj.doh, sirius0323, taejunham, jaewlee, gajh\}@snu.ac.kr
\end{tabular}
}

\begin{document}
\maketitle
\thispagestyle{plain}
\pagestyle{plain}


\begin{abstract}
Since its public introduction in the mid-2010s, the Row Hammer (RH) phenomenon has drawn significant attention from the research community due to its security implications. Although many RH-protection schemes have been proposed by processor vendors, DRAM manufacturers, and academia, they still have shortcomings.
Solutions implemented in the memory controller (MC) incur increasingly higher costs due to their conservative design for the worst case in terms of the number of DRAM banks and RH threshold to support.
Meanwhile, DRAM-side implementation either has a limited time margin for RH-protection measures or requires extensive modifications to the standard DRAM interface.
Recently, a new command for RH-protection has been introduced in the DDR5/LPDDR5 standards, referred to as \emph{refresh management (RFM)}.
RFM enables the separation of the tasks for RH-protection to both MC and DRAM by having the former generate an RFM command at a specific activation frequency and the latter take proper RH-protection measures within a given time window.
Although promising, no existing study presents and analyzes RFM-based solutions for RH-protection.
In this paper, we propose \name, the \emph{first} RFM interface-compatible, DRAM-MC cooperative RH-protection scheme providing deterministic protection guarantees. \name has minimal energy overheads for common use cases without adversarial memory access patterns.
We also introduce \nameplus, an optional extension to provide minimal performance overheads at the expense of a tiny modification to the MC, while utilizing existing DRAM commands.
\end{abstract}

\section{Introduction}
\label{sec:introduction}

\emph{Row Hammer} (RH) has been critical DRAM reliability and security vulnerabilities that have troubled the industry for almost a decade. This refers to a phenomenon in which a certain frequently activated row (aggressor) results in bit-flips in the corresponding adjacent rows (victims). In particular, RH is incurred when the activation rate exceeds the RH threshold ($\nth$). RH is especially dangerous as it breaks the basic integrity guarantee in the computer system and can be abused in various attack scenarios\mbox{~\cite{exploit_antidos,pc_attack1,server_attack1,server_attack2,server_attack3,mobile_attack1,micro-2020-pthammer}.
}

The criticality of this problem has motivated many RH-protection solutions. There exist several software-based solutions~\cite{anvil,page-allocation-method1,mobile_attack1,mascat, ZebRAM}, but such of these typically incurs a high-performance cost and have limited coverage (i.e., only effective against a specific attack scenario). For these reasons, architectural solutions have emerged as promising alternatives.

One of the important design decisions for an architectural RH-protection scheme is to determine where to implement the proposed solution within the system. In practice, most RH-protection solutions are either implemented in an on-die memory controller (MC) or a DRAM device. For example, Graphene~\cite{micro-2020-graphene}, BlockHammer~\cite{blockhammer}, and PARA~\cite{PARA} have been proposed for implementation on the processor-side MC, whereas TWiCe~\cite{TWiCe} and industry-oriented RH-protection schemes~\cite{host_assisted_patent, trrespass} are implemented in DRAM. Unfortunately, both choices have their own drawbacks. 

First, the MC-side implementation needs to provision RH-protection resources for the \emph{worst-case} scenario, where the expected $\nth$ level is very low and the processor is connected to the maximum number of DRAM banks it supports. As a result, this strategy tends to require a large extra area for the counter structures utilized by the RH-protection mechanism. DRAM-side implementations are free from such concerns, as $\nth$ of a specific DRAM is more accurately estimated by DRAM vendors, and the resource usage is proportional to the number of DRAM banks because on-DRAM RH-protection schemes are often deployed on a per-bank or per-DIMM basis. However, such on-DRAM protection schemes have \emph{interface} issues. To secure the time margin for the extra operations for potential RH victim rows, DRAM-side schemes must either request the MC to generate non-standard adjacent row refresh (ARR) commands or perform extra operations during the auto-refresh process (ordinary DRAM operation) in a way transparent to the MC. The former mechanism breaks the abstraction that DRAM is a passive device, whereas the latter~\cite{trrespass}, referred to as the time-margin-stealing method, is not always possible depending on DRAM characteristics such as the time margin during the auto-refresh process.

\emph{Refresh Management (RFM)} is a newly added extension for the latest DDR5 and LPDDR5 interfaces\mbox{~\cite{jedec-ddr5,JEDEC_LPDDR5}}, allowing the DRAM-side implementation of an RH-protection solution to cooperate smoothly with an MC.
An MC sends an RFM command at a specific activation frequency to a target DRAM bank without specifying a target row.
The DRAM-side RH-protection scheme exploits the time margin provided by the RFM command to undertake necessary operations.
This cooperation between the MC and DRAM effectively avoids the critical drawbacks of MC- or DRAM-side only implementations.

\begin{table*}[tb!]
\centering
\caption{Categorization of existing Row Hammer mitigation schemes and Mithril}
\vspace{-0.05in}
\label{tab:3_remedy_and_algorithm}
\scalebox{0.95}[0.95]{
\begin{tabular}{l|l|l|l|l} 
\Xhline{3\arrayrulewidth}
\hline
 \textbf{Mitigation Scheme}
       & \begin{tabular}[c]{@{}c@{}}\textbf{Protection Guarantee}\end{tabular}
       & \begin{tabular}[c]{@{}c@{}}\textbf{Remedy}\end{tabular}
       & \begin{tabular}[c]{@{}c@{}}\textbf{Implementation Location}\end{tabular}
       & \begin{tabular}[c]{@{}c@{}}\textbf{Tracking Mechanism}\end{tabular} \\ 
\hline 
 \textbf{PARA~\cite{PARA}}
       & Probabilistic
       & ARR
       & MC
       & Probabilistic Approach \\ 
\hline
 \textbf{CBT~\cite{isca-2017-cbt,ieee-cal-2017-cbt}}
       & Deterministic
       & ARR
       & MC
       & Grouped Counter Approach \\
\hline 
 \textbf{TWiCe~\cite{TWiCe,ieee-cal-2018-twice}}
       & Deterministic
       & ARR (feedback-augmented)
       & DRAM (buffer-chip)
       & Streaming Algo. (Lossy-Counting) \\
\hline
 \textbf{Graphene~\cite{micro-2020-graphene}}
       & Deterministic
       & ARR
       & MC
       & Streaming Algo. (Counter-based Summary) \\
\hline
 \textbf{BlockHammer~\cite{blockhammer}}
       & Deterministic
       & Throttling
       & MC
       & Streaming Algo. (Count-min Sketch) \\
\hline
 \textbf{Mithril}
       & Deterministic
       & RFM
       & DRAM (co-op with MC)
       & Streaming Algo. (Counter-based Summary) \\
\Xhline{3\arrayrulewidth}
\end{tabular}
}
\vspace{-0.05in}
\end{table*}

Despite its promising traits, the applicability of RFM as an RH-protection scheme has not been publicly verified or properly evaluated to the best of our knowledge.
A prior probabilistic scheme~\cite{PARA} can be trivially applied. 
However, prior deterministic (guaranteeing not to exceed $\nth$) schemes cannot be directly applied to the RFM interface.
Prior ARR-based schemes \emph{reactively} issue a command targeting a specific row when the activation count reaches a scheme-specific \emph{predefined threshold}.
However, given its periodicity, the RFM interface is prone to the worst-case scenario where a large number of rows will simultaneously require a preventive refresh in a short time period, unlike the ARR-based schemes.
Thus, prior approaches are not compatible with the RFM interface.

In this paper, we propose \name, a novel RFM-compatible, deterministic RH-protection scheme that exploits MC and DRAM in a cooperative manner.
To avoid the aforementioned concentration of rows to refresh for RH-protection, we utilize a \emph{greedy approach} when selecting the target row to refresh upon every RFM command.
We investigate the effective use of streaming algorithms\mbox{~\cite{streaming_algo_textbook}} (Section~\ref{sec:contribution-1}) and provide a new mathematical proof through which we guarantee deterministic protection by maintaining the greedy selection scheme (Section~\ref{sec:contribution-2} and Appendix).

Finally, we propose 1) a hardware scheme to obviate the need for counter table resets, which were mandatory in prior studies; 2) an algorithmic optimization for energy savings; and 3) an extension to the RFM interface to mitigate the performance overhead by exploiting the memory access patterns of ordinary workloads.

The key contributions of this paper are as follows:
\setlist{nolistsep}
\begin{itemize}[noitemsep,leftmargin=0.2in]
  \item We propose \name, the first RFM-based RH-protection scheme with deterministic safety guarantees, exploiting a modified Counter-based Summary algorithm~\cite{misra_gries, Space_Saving}.
  \item We provide a rigorous mathematical proof of the modified algorithm and the RH safety of \name.
  \item We suggest energy and performance optimization techniques that exploit the memory access patterns of common, non-adversarial workloads.
\end{itemize}

\section{Background}
\label{sec:background}

\subsection{DRAM Refresh}
\label{subsec:dramrefresh}
DRAM stores a single bit in a \emph{cell}, composed of one capacitor and one access transistor~\cite{isca-2014-rbd}.
These cells are organized into \emph{rows} and \emph{columns}.
A DRAM row, the cells of which share a wordline, is the granularity of the \emph{activation} (ACT) and \emph{precharge} (PRE), respectively allowing and disallowing \emph{read} or \emph{write} operations on the row.
The read and write operation involves accessing a certain number of columns in an activated row.
DRAM is composed of multiple \emph{banks}.
Each bank allows independent ACT, PRE, read, and write operations. 
Multiple banks form a \emph{rank}, which shares the memory channel with other ranks and the memory controller (MC) at the host side.

Due to the inherent characteristic of a DRAM cell capacitor, by which the stored charge leaks over time, the cell value must be restored periodically~\cite{DRAM_refresh_tradeoff, synlec-2019-memory}.
This type of periodic restoration, referred to as an \emph{auto-refresh}, is initiated at every refresh (REF) command within the tRFC (refresh time) period.
Every DRAM row must be refreshed at least once during every refresh window period (tREFW) to be safe from this charge retention problem.
In modern DRAM devices (e.g., DDR5~\cite{jedec-ddr5}), all rows in a single bank are divided into typically 8,192 groups.
A group is refreshed in every time interval tREFI (refresh interval).

\subsection{Row Hammer Phenomenon}
\label{subsec:rowhammer}
Row Hammer (RH) refers to a phenomenon in which repetitive activations of a specific row (aggressor) lead to bit flips in physically nearby rows (victims)~\cite{PARA-Retrospective, PARA, Row_Hammer_Mechanism, threshold_scaling}.
A bit flip is observable when the ACT count reaches a certain RH threshold ($\nth$) without being refreshed inside a tREFW time window. Because two aggressors can simultaneously affect a single victim, $\sfrac{\nth}{2}$ ACTs on each aggressor can cause a bit flip (double-sided attack).
The $\nth$ value varies depending on different chips, generations, and/or DRAM manufacturers~\cite{revisiting}.
The RH problem has worsened following the current scale-down trend of fabrication technology, due to the intensified inter-cell interference.
Recent studies~\cite{trrespass, revisiting} reported that $\nth$ has been reduced to a mere several thousand ACTs.
It has also been observed that non-adjacent rows affect the victim rows when activated frequently, which degrades the effective $\nth$.

\subsection{Classifying Prior RH Mitigation Schemes}
\label{sec:prior_mitigating_schemes}
As shown in Table~\ref{tab:3_remedy_and_algorithm}, existing architectural RH-protection schemes all have four important criteria of a 1) protection guarantee, 2) type of remedy, 3) implementation location, and 4) tracking mechanism.

\subsubsection{Protection Guarantee}
There exist two different types of RH-protection guarantees, deterministic and probabilistic.
The \textbf{deterministic guarantee} ensures RH-protection by guaranteeing that a victim row is always refreshed before the number of ACTs exceeds $\nth$ on its aggressors, either by an extra preventive refresh or the normal auto-refresh.
This type utilizes a counter structure to track the aggressor row and deals with it by applying a certain remedy.
The main drawback of a deterministic scheme is its higher area overhead due to the large counter structure.

The \textbf{probabilistic guarantee} prevents RH with a certain probability. 
The probabilistic approach has its strength in the minimal area overhead.
However, the performance overhead is exacerbated severely when the target $\nth$ level is lowered or when the number of DRAM devices in the system increases.
It does not provide a deterministic protection guarantee, either.

\subsubsection{Remedies of Prior RH-protection Schemes}
Prior works exploited one of two remedies, adjacent row refresh (ARR) or throttling.
\textbf{ARR} refers to a type of command that the MC issues to DRAM with an explicit \emph{target row address} (either aggressor or victim) at a \emph{required moment}.
It triggers an extra preventive refresh on the potential RH victim rows within the time margin provided by the command.
This differs from the normal REF command, which is row-agnostic and periodic.
Prior RH-protection schemes that exploited ARR either issued commands based on some \emph{probability}~\cite{PARA, PRoHIT, MRLoc} or when the ACT count of a certain aggressor exceeds a scheme-specific \emph{predefined threshold}, which is assumed to be hazardous.
However, ARR is not practically applicable because it either requires a new interface that breaks the abstraction of a passive DRAM device or requires the MC to become the sole manager of RH-protection. In fact, a command with a similar concept was once proposed in DDR4, but is now deprecated.

\textbf{Throttling} is a method by which the MC \emph{delays} the frequency of activation on an aggressor starting at the moment of identification for a defined time.
The duration and intensity of the delay are adjusted to guarantee RH-protection.
After the initial suggestion of such methodology~\cite{throttling_patent}, a deterministic RH-protection scheme utilizing the throttling method was proposed~\cite{blockhammer}.
However, leveraging throttling requires system-level support along with more complex MC scheduling and makes the system vulnerable to adversarial patterns (details in Section~\ref{sec:6_comparison_compatible}).

\subsubsection{Implementation Location}
Prior RH-protection schemes are all located either on the MC-side or the DRAM-side.
\textbf{MC-side implementation} has strength in that it utilizes a superior logic process with a larger area budget.
However, it has the following major drawbacks.
First, it requires a conservatively high number of counter structures to populate.
The counter table of the deterministic scheme is typically allocated per DRAM bank.
The latest CPU servers, such as Intel Ice Lake, support up to 1,024 banks per socket (8 channels $\times$ 8 ranks $\times$ 16 banks).
This number could increase further if we consider 3D stacked DRAM devices or future generations.
Despite the fact that fully populating 1,024 banks may be unlikely, the counter structures must be designed to support the worst case.
Second, MC-side implementation must protect against a conservatively low target $\nth$ value.
The target $\nth$ varies greatly depending on the manufacturer, generation, or even the device.
Considering that most deterministic schemes must be tuned to the target $\nth$ at the time of their design, they must 
protect against pessimistic $\nth$ values.

\textbf{DRAM-side implementation} typically relies on an extra preventive refresh on a potential RH victim row.
However, it is difficult to secure adequate uninterrupted time to execute preventive refreshes in the conventional MC-DRAM interface.
Previous DRAM-side RH-protection schemes attempted to address this problem with either a \emph{feedback-augmented} ARR command\mbox{~\cite{TWiCe}} or via the auto-refresh time-margin stealing method\mbox{~\cite{trrespass}}.
The former is similar to the normal ARR command issued by MCs but requires that DRAM halt the MC for a certain amount of time.
There exist some methods of feedback from DRAM to MC, such as an ALERT\_n signal, but these require more pins to deliver additional alert types to support the DRAM-side RH-protection scheme.
The latter method, auto-refresh time-margin stealing, invisibly executes a preventive refresh during the normal auto-refresh operation.
Although not requiring any feedback path, it has a limited time margin that can be stolen and thus cannot be scaled to a low $\nth$ value.

\begin{figure}[tb!]
    \begin{center}
    \includegraphics[width=0.9\columnwidth]{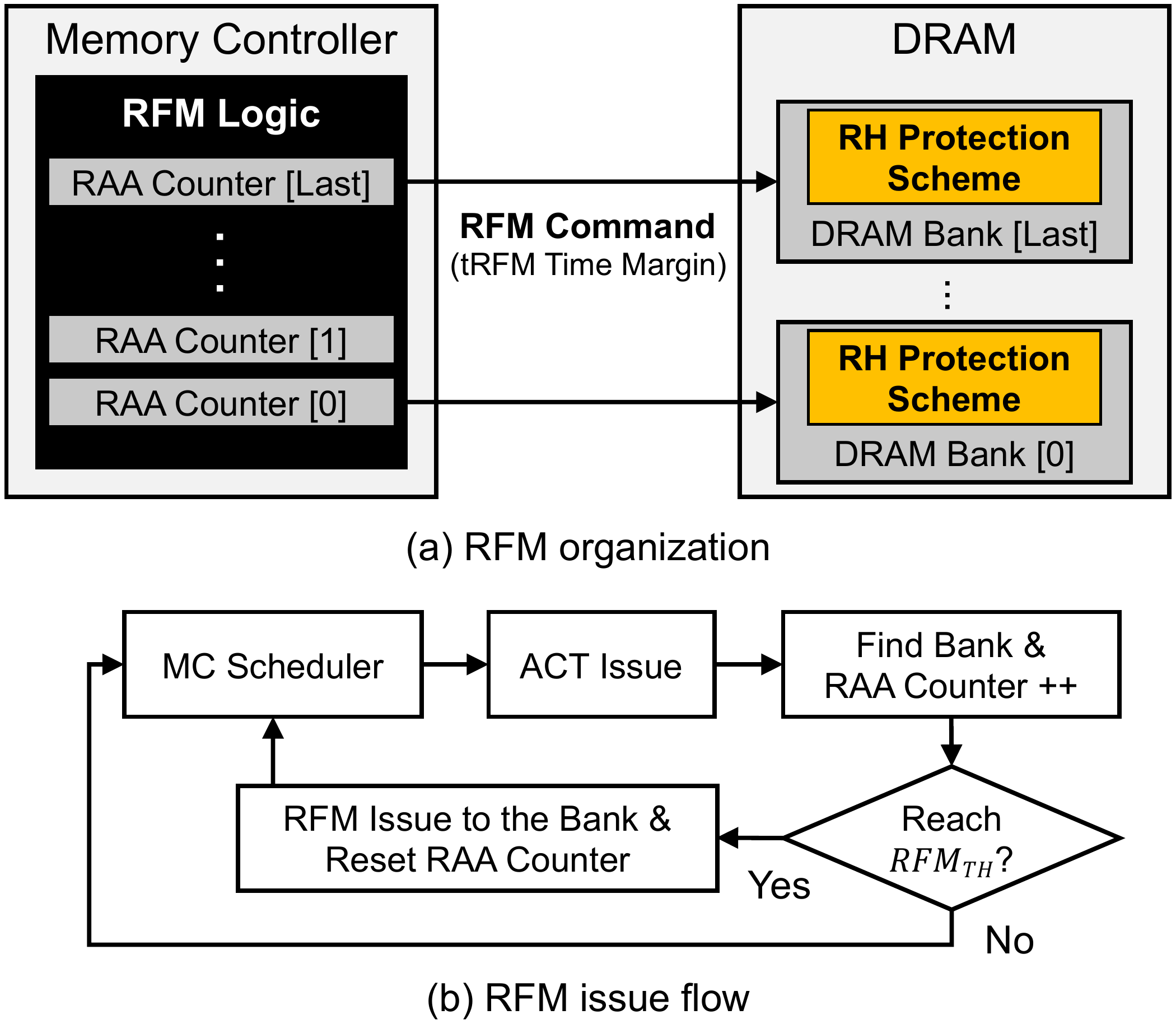}
    \end{center}
    \vspace{-0.12in}
    \caption{(a) Example of main-memory organization with the RFM interface-, and (b) the issue logic of RFM.}
    \vspace{-0.12in}
    \label{fig:2_back_RFM_HW}
\end{figure}

\begin{table*}[!tb]
    \centering
    \small
    \caption{Symbols and their descriptions used for DRAM refresh, RH, and RFM}
    \begin{tabular}{p{0.27\columnwidth}p{0.825\textwidth}}
        \Xhline{3\arrayrulewidth}
        \textbf{Symbol} & \textbf{Description} \\
        \Xhline{1.5\arrayrulewidth}
        tREFW & Per row auto-refresh interval (e.g., 32ms or 64ms) \\ 
        $\nth$ & RH threshold \\
        $\rfmth$ & RFM threshold \\
        Preventive refresh & Extra refresh of potential RH victim rows. Executed during ARR, RFM command, or hidden under auto-refresh. \\
        \Xhline{0.5\arrayrulewidth}
        \Xhline{3\arrayrulewidth}
    \end{tabular}
    \label{tbl:2_back_notation}
    \vspace{-0.1in}
\end{table*}

\subsubsection{Tracking Mechanism and Streaming Algorithms}

Each RH-protection scheme has its own tracking mechanism to identify the aggressor or victim rows with high ACT counts.
The tracking mechanism of a probabilistic scheme is often insignificant. However, for a deterministic scheme, it is crucial to choose an effective tracking mechanism to minimize the area overhead of the counter structure. One class of tracking mechanisms is based on \emph{streaming algorithms}\mbox{~\cite{streaming_algo_textbook}}, which are most effective when estimating the ACT counts of rows when the counter table size is limited. Multiple prior works\mbox{~\cite{TWiCe, micro-2020-graphene, blockhammer}} explicitly leverage or can be interpreted as based on such streaming algorithms.

The streaming algorithm was first invented and developed unrelated to the RH problem in the field of data mining to analyze fast and dense data streams with limited memory. A certain subset of the algorithms estimates the total number of occurrences per input element. Considering the fact the ACT commands with an address are ``streamed'' from the MC to DRAM, a subset of the streaming algorithms can be utilized to estimate the ACT count per row address. Thus, they are suitable as an effective tracking mechanism of an RH-protection scheme.
They report the approximate number of occurrences for each element (address), referred to as the \emph{estimated count}, instead of the \emph{actual count}. Generally, the resolution (or the error) of the algorithm is higher (lower) when more memory is used.

Several other works\mbox{~\cite{isca-2017-cbt,ieee-cal-2017-cbt, cat-two}} use the different approach of a grouped counter. They allocate multiple rows to a single counter to reduce the area overhead of the tracking mechanism.
They optimize further by dynamically adjusting the allocation or by utilizing the characteristics of DRAM.

\subsection{RFM Interface as a New Remedy}
\label{sec:rfm_command}

The RFM interface has been newly introduced as an \emph{alternative remedy} that allows for \emph{DRAM-MC cooperation}.
It is suggested as the primary means of RH-protection by the JEDEC committee~\cite{jedec-near_term,jedec-system-level}.
The RH-protection scheme resides on the DRAM-side while the MC provides a periodic but DRAM-row agnostic time margin to the DRAM bank.
Periodic here is not based on time but on the number of ACTs over a single DRAM bank.
Figure~\ref{fig:2_back_RFM_HW} shows an example of a main-memory organization scheme using an RFM interface and RFM issue logic.
An MC has a Rolling Accumulated ACT (RAA) counter per bank that keeps track of the number of ACTs on its bank.
When the RAA count reaches the RFM threshold ($\rfmth$) set by the DRAM device, the MC issues an RFM command only to the corresponding bank and resets the RAA counter for the target bank.
The larger the $\rfmth$, the lower the frequency of the RFM command, which reduces the effect on the system performance.
At every RFM command issue, the recipient bank receives a time margin (tRFM) during which no disturbance from any other regular operation is guaranteed.

A key difference with regard to the prior ARR command is that RFM is row agnostic and periodic (i.e., it cannot be issued in a bursty way).
In a sense, it can be seen as an extension of the time-margin stealing method. The format of an RFM command is similar to that of a per-bank REF command~\cite{jedec-ddr5, JEDEC_LPDDR5} specifying the bank to apply RFM, but not a certain row. Therefore, it requires minimal additional complexity to the MC.
The symbols related to DRAM refresh, RH, and RFM are summarized in Table~\ref{tbl:2_back_notation}.

\section{Investigating RFM-based Schemes}
\label{sec:contribution-1}

RFM as a remedy for RH-protection allows for DRAM-side implementation with MC cooperation, eliminating multiple drawbacks of MC-side- or DRAM-side-only implementation.
First, RFM can minimize the aforementioned overkill of the MC-side-only implementation because it can use an accurate prediction of $\nth$ and even set the $\rfmth$ value after testing the manufactured DRAM chip.
It also scales according to the number of DRAM devices that are actually attached to the host. 
Second, RFM also provides a standard interface that a DRAM-side RH-protection scheme can utilize to gain an additional time margin for RH preventive refreshes. The ARR command assumed in many prior works is not supported in the recent DDR interface. RFM is newly being adopted and is now recommended as the primary method for RH-protection\mbox{~\cite{jedec-system-level,jedec-near_term}}.

\subsection{Incompatibility of Prior Approaches}
\label{sec:incompatibility_of_prior_works}

\begin{figure}[tb!]
    \begin{center}
    \includegraphics[width=1\columnwidth]{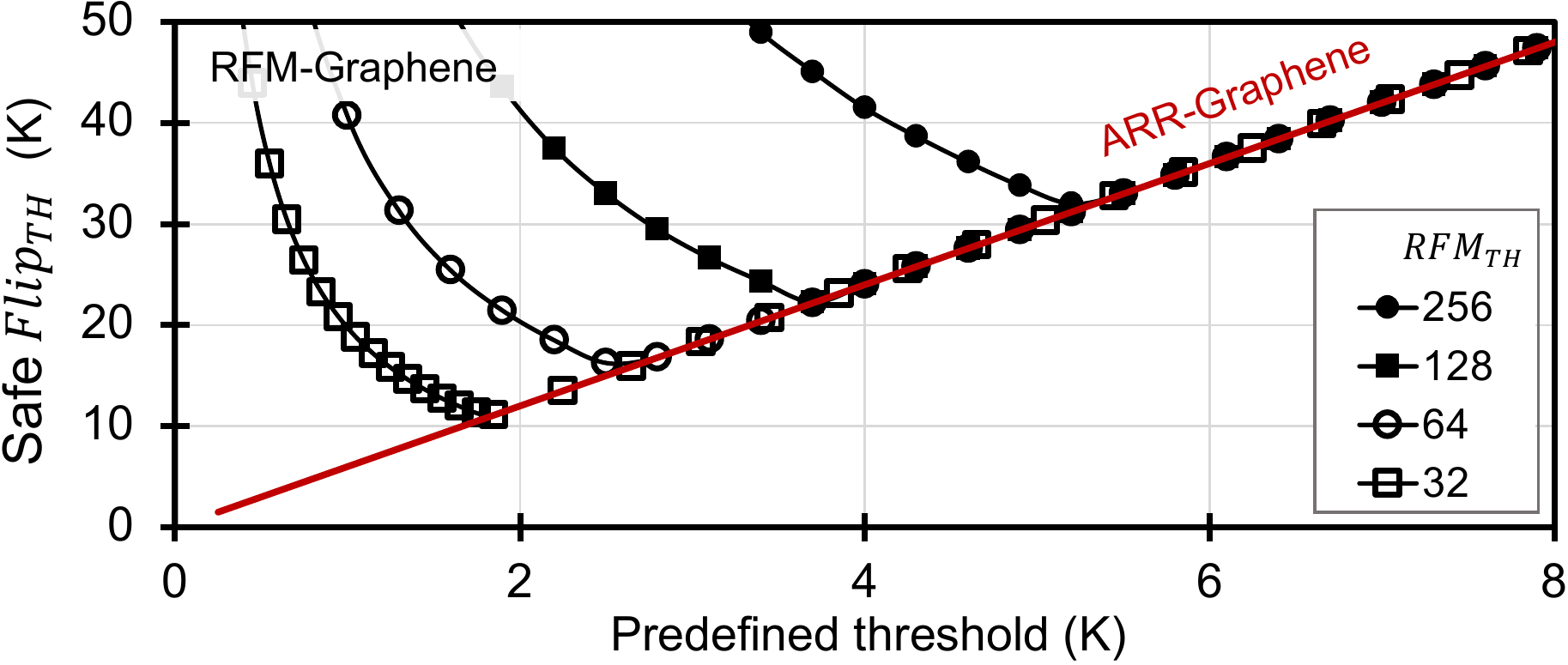}
    \end{center}
    \vspace{-0.15in}
    \caption{Ineffectiveness of RFM-Graphene compared to the original ARR-Graphene. Note that the inverse of the predefined threshold indicates a larger table size with a lower resolution.}
    \vspace{-0.1in}
    \label{fig:3_ARR-based_RFM}
\end{figure}

Although promising, prior approaches based on ARR are not effective in RFM because RFM is vulnerable to the concentration of victim rows that require a preventive refresh. 
The ARR-based scheme has its own predefined threshold value directly related to the target $\nth$.
When its tracking mechanism detects the ACT count of an aggressor row reaching the predefined threshold, it immediately issues an ARR command and executes preventive refreshes to guarantee the deterministic RH safety.
For example, Graphene with ARR can provide safety for $\nth$ that is linear to the predefined threshold (red line in Figure~\ref{fig:3_ARR-based_RFM}).
Even if the predefined threshold is low, the relationship between predefined threshold and $\nth$ does not change.

However, when this ARR-based approach is applied to the RFM interface, there is a limit to $\nth$ that is guaranteed to be safe regardless of how low the predefined threshold is set (see Figure~\ref{fig:3_ARR-based_RFM}).
With the same prior approach, one scheme could set a predefined threshold and buffer the aggressor rows that reach it.
Then, when the subsequent RFM command is issued, the postponed preventive refresh can be executed on the corresponding adjacent victim rows.
However, such a scheme is vulnerable when multiple aggressor rows reach the predefined threshold in a short period.
For example, when the predefined threshold is 2K and the $\rfmth$ is reasonably set to 64 (see Section~\ref{sec:analysis}), the safe $\nth$ becomes 20K, not 10K.
This occurs because 310 rows can reach 2K in a single tREFW period; thus the last buffered row must wait through (310$\times$64) ACTs.

\subsection{Greedy Selection}
\label{sec:sub_greedy_selection}

To prevent the concentration of victim rows requiring a preventive refresh in an RFM-based scheme, it is necessary to properly select the target row and refresh its victims, even if the ACT count of the row has not reached $\nth$ or another predefined threshold.
In particular, \emph{we propose the use of the greedy selection of a target row upon every RFM command for the RFM-based scheme.}

An intuitive method for the proper selection of a row at every RFM command is to greedily choose the row with the highest estimated ACT count based on the tracking mechanism. Also, after choosing the row and refreshing its victims, it is logical to reset or minimize the estimated ACT count of the selected row to assist with the decision at the next RFM command, as the actual ACT count is now $0$ after the refresh. Based on this simple basic principle, we search for the proper tracking mechanism.

\subsection{Counter-based Summary}
\begin{figure}[tb!]
    \begin{center}
    \includegraphics[width=0.95\columnwidth]{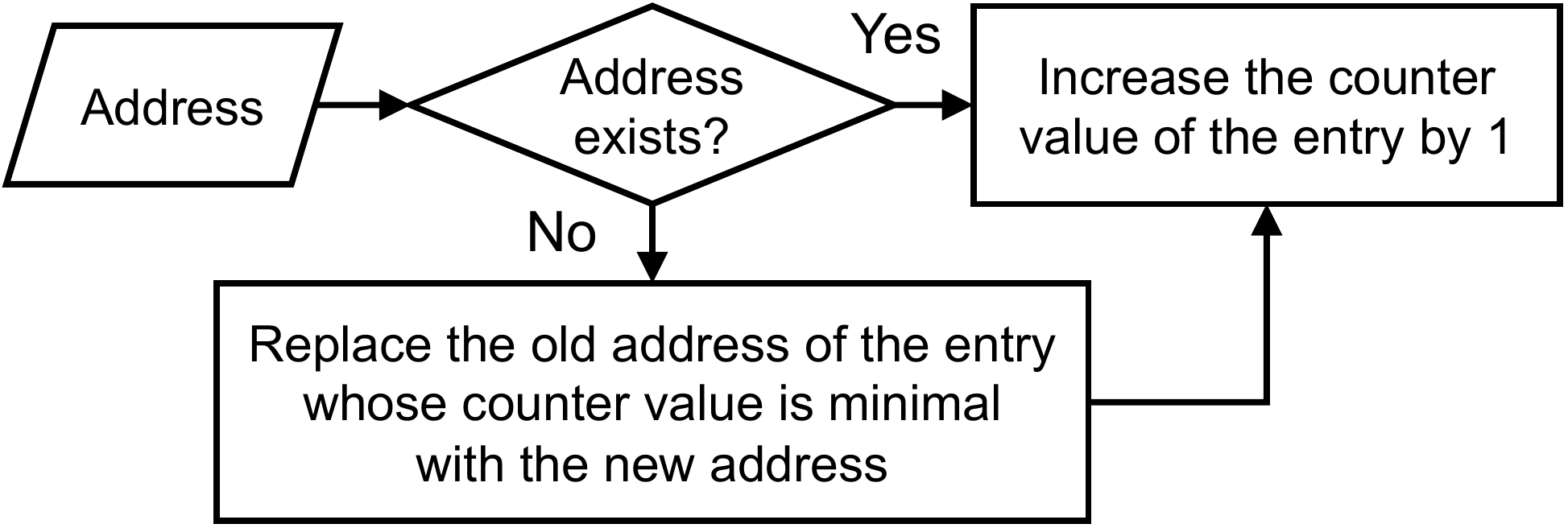}
    \end{center}
    \vspace{-0.15in}
    \caption{Counter-based Summary (CbS) algorithm operation.}
    \vspace{-0.1in}
    \label{fig:2_back_space_saving}
\end{figure}

\label{sec:3_counter_based_summary}

We choose to use some variant of streaming algorithms for the RFM-based RH-protection scheme. While the grouped counter approach was effective in ARR-based work, it is no longer efficient in RFM (Section~\ref{sec:3_grouped_counter}). To support the greedy selection policy properly, the streaming algorithm must link the actual ACT count to the lower and upper bound of the estimated ACT count. We explain this in detail with an example.

\emph{Counter-based Summary} (CbS) algorithm\mbox{~\cite{misra_gries, Space_Saving, mergeable_summary}} is a representative streaming algorithm that matches such needs. The CbS algorithm has a table of entries, each holding an address and a counter.
When the queried address hits an entry in the table (on-table), the counter in the corresponding entry is incremented by one.
When it misses the table (off-table), it replaces the address of the entry with the minimum counter value in the table with the queried address.
It then increments its counter by one (see Figure~\ref{fig:2_back_space_saving}).
Due to its monotonically increasing nature and swapping, the accumulated counter value above the minimum in the table belongs to the currently written address.
In contrast, the ones below the minimum cannot find their source.

\Scale[0.85]{
\begin{minipage}{1.0\columnwidth}
\vspace{0in}
\begin{align*} 
\text{On-Table\ Addr:}\  Estimated\ Count &= Written\ Counter\ Value\\
\vspace{-0.18in}
\text{Off-Table\ Addr:}\  Estimated\ Count &= Min \\
 Actual\ Count &\leq Estimated\ Count\,\,\;\;\;\;\;\;(1) \\
 Estimated\ Count &\leq Actual\ Count + Min\;\;(2)
\end{align*}
\end{minipage}}\vspace{0.075in}\\

The CbS algorithm reports the estimated (ACT) count of an on-table address with its written counter value, whereas the count of an off-table address is estimated with the minimum value in the entire table. Inequalities (1) and (2) correspondingly 
show the lower-bound and upper-bound of the estimated count in relation to the actual (ACT) count. \textit{Min} denotes the minimum counter value in the table.

First, based on the lower bound (inequality (1)) of the estimated count, the RH-protection scheme is able to act upon an inaccurate, yet conservatively large ACT value. This allows the scheme to provide deterministic safety\mbox{~\cite{TWiCe, micro-2020-graphene, blockhammer}}. Second, the upper-bound (inequality (2)) of the estimated count is also necessary to decrement the estimated count of the greedily selected row at the RFM command, where the actual ACT count is now $0$. Without this upper-bound, the estimated count cannot be decremented safely. The lossy-counting algorithm used in TWiCe\mbox{~\cite{TWiCe}} also has both the lower and upper bound of the estimated counts, but is less efficient algorithmically (as is later shown in Figure~\ref{fig:4_nentry_rfmth}). It causes fewer preventive refreshes at the cost of a higher area overhead. Thus, we choose the CbS algorithm as the basic building block of our tracking mechanism.

There exists other streaming algorithms that only have a lower bound of the estimated count, such as Count-min Sketch\mbox{~\cite{Count_Sketch}}, but it can only be used in throttling based works such as BlockHammer\mbox{~\cite{blockhammer}}. Others that do not have the lower bound such as Sticky-sampling\mbox{~\cite{Sticky_Sampling_and_Lossy_Counting}} or Count-sketch\mbox{~\cite{Count_Sketch}} cannot provide deterministic safety.

\begin{figure}[tb!]
    \begin{center}
    \includegraphics[width=0.9\columnwidth]{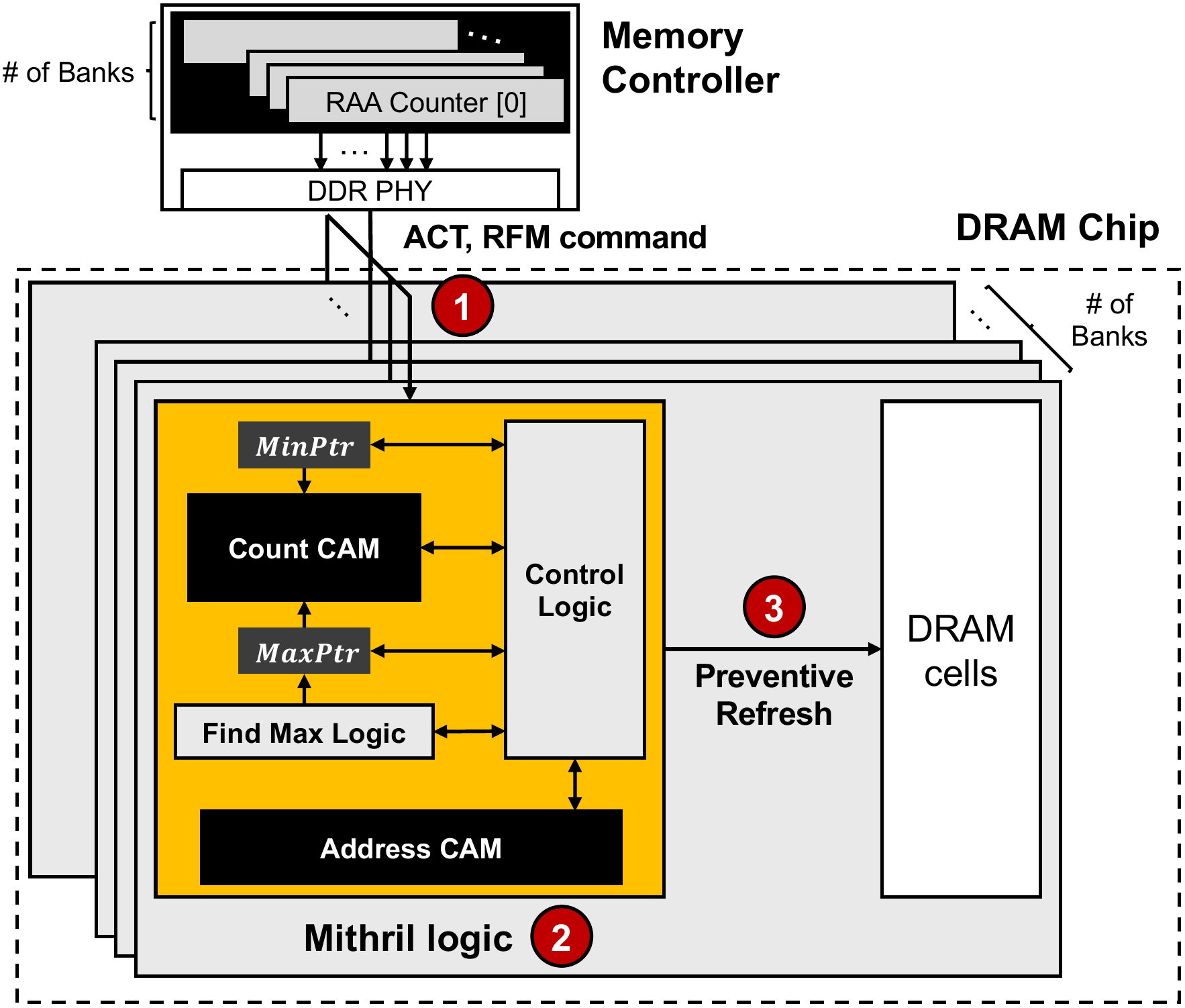}
    \end{center}
    \vspace{-0.1in}
    \caption{Mithril hardware implementation. An identical Mithril module with the logic and CAM structure is populated per bank, at every DRAM chip. Here, \textcircled{1}, \textcircled{2}, and \textcircled{3} denote the high-level command flow.}
    \label{fig:mithril_hardware_implementation}
    \vspace{-0.1in}
\end{figure}

\subsection{Grouped Counter Approach}
\label{sec:3_grouped_counter}

The grouped counter approach was another type of tracking mechanism in ARR-based works.
However, prior works that augmented this methodology are not compatible with or efficient at the RFM interface.
CBT~\cite{isca-2017-cbt,ieee-cal-2017-cbt} is the representative scheme of this type.
First, it cannot utilize the RFM opportunities during its tree construction phase.
Suppose it chooses to refresh a group prematurely that is not fully split. In such a case, it will have to refresh many rows too conservatively. 
Second, even after the tree is constructed, having a leaf node of a size larger than eight rows will not fit into a single tRFM period, leading to the stacking of refresh loads.
CAT-TWO~\cite{cat-two}, which extends CBT, may guarantee that a leaf is small (covering a single row) enough, but only at the cost of a higher area overhead.

\begin{figure*}[tb!]
    \vspace{-0.1in}
    \begin{center}
    \includegraphics[width=0.97\textwidth]{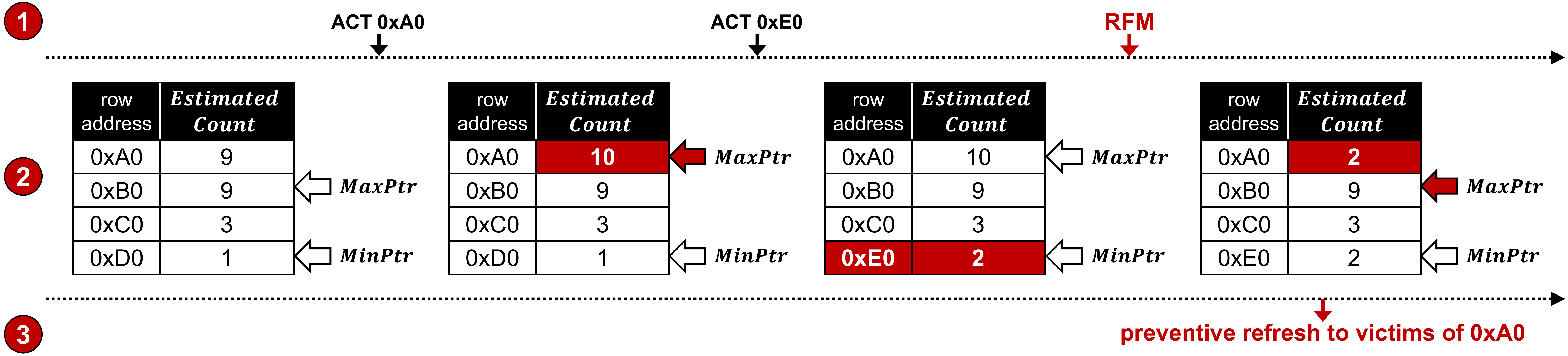}
    \end{center}
    \vspace{-0.1in}
    \caption{
    Sequence of ACT and RFM commands and the corresponding update of the \name table.  \textcircled{1}, \textcircled{2}, and \textcircled{3} correspond to those in Figure~\ref{fig:mithril_hardware_implementation}.}
    \label{fig:4_cont2_mithril_operation_example}
    \vspace{-0.1in}
\end{figure*}

\subsection{Probabilistic RFM-based Scheme}
\label{sec:parafm}

An RFM-compatible probabilistic RH-protection scheme (henceforth PARFM) can be built in a manner similar to PARA\mbox{~\cite{PARA}}. Whenever an RFM command arrives, PARFM randomly samples a single aggressor row among the last $\rfmth$ ACTs. PARFM's protection capability depends solely on $\rfmth$. By adjusting $\rfmth$ properly, PARFM can guarantee probabilistic safety on the target $\nth$.
However, as $\nth$ decreases, PARFM requires a lower $\rfmth$ than those in deterministic RFM-based schemes to maintain a high safety probability, leading to greater performance and energy overhead.
We discuss this further in Section~\ref{sec:analysis}.

\section{\name}
\label{sec:contribution-2}
Based on the investigation of the RFM-based RH-protection schemes in Section~\ref{sec:contribution-1}, we present \name, the first RFM-interface-compatible RH-protection scheme providing a deterministic protection guarantee.
It exploits a modified CbS algorithm for counter management.

\subsection{Organization}
\label{sec:4_organization}

The Mithril logic in each DRAM bank is composed of a counter structure (henceforth the Mithril table), two pointers ($\maxptr$ and $\minptr$), and the control logic (Figure~\ref{fig:mithril_hardware_implementation}). To be more specific, the Mithril table comprises two CAM structures, one storing the row address and the other the ACT count. Each ACT counter is directly related to a single row address. The $\maxptr$ and $\minptr$ pointers are also employed as index pointing registers. The Mithril structure including the CAMs and logic must be equipped in every bank at every DRAM chip (Figure~\ref{fig:mithril_hardware_implementation}).

\subsection{Operation}
\label{sec:4_operation}

Figure~\ref{fig:4_cont2_mithril_operation_example} illustrates how Mithril manages the corresponding Mithril table and the two pointers, $\maxptr$ and $\minptr$.
The Mithril logic of the corresponding DRAM bank is informed at every ACT command (with an address) or RFM command (without an address).
If the Mithril logic receives an ACT command, the count CAM, $\maxptr$, and $\minptr$ are updated. To be more specific, first, Mithril checks if the address table already tracks the activated row address.
If so, the associated ACT counter is incremented by one. When the row address misses, the address of the entry indicated by $\minptr$ is replaced with the requesting row address, and its counter is incremented by one. If affected, $\maxptr$ and $\minptr$ are updated at each step to point correspondingly to the correct maximum and minimum. Thus far, the operation is identical to that of the original CbS algorithm.

When the Mithril logic instead receives an RFM command, Mithril selects the entry pointed via $\maxptr$ (greedy-selection).
It performs a preventive refresh for the two victim rows associated with this entry, identified as the prime candidates of the aggressor rows.
Then, the counter value is decremented to the table's minimum value pointed by $\minptr$. $\maxptr$ is also updated correspondingly. The new $\maxptr$ must be found during the RFM time window.

\subsection{Mathematical Proof of Protection Guarantee}
\label{sec:brief_overview_proof}
Mithril guarantees RH safety by preventing the ACT count of any row from reaching $\nth$ by continuing the greedy selection and preventive refresh processes.
This contradicts prior works which triggered a preventive refresh at the exact hazardous moment where a row reaches a predefined threshold ACT value.
To prove the deterministic safety of Mithril, we initially prove that \emph{continuously applying greedy selection and preventive refresh processes creates an upper bound in the rate of the estimated ACT count increment during tREFW.}
That upper bound is defined by an equation with $\entrynum$ (the number of Mithril counter entries) and $\rfmth$, as follows:

\vspace{0.1in}
\noindent\textbf{Theorem 1.} \emph{Within any tREFW, an increase in the estimated count for any single row is bounded to $\maxcount$, which is a function of $\entrynum$ and $\rfmth$.}\\
\vspace{-0.1in}

\Scale[0.8]{
\begin{minipage}{1.0\columnwidth}
\begin{align*}
M = \sum_{k=1}^{\entrynum} \frac{\rfmth}{k} + \frac{\rfmth}{\entrynum} \left(\frac{\text{tREFW}(1-\frac{\text{tRFC}}{\text{tREFI}})}{\text{tRC} \times \rfmth+\text{tRFM}} - 2\right)
\end{align*}
\end{minipage}}\vspace{0.075in}\\

\noindent Then, by setting $\entrynum$ and $\rfmth$ so that $M$ is less than ($\nth$/2), Mithril can deterministically prevent RH from experiencing double-sided attacks.
The detailed proof of Theorem 1 is provided in the Appendix (Section~\ref{sec:appendix}).

\subsection{Configuring $\entrynum$ and $\rfmth$} 
\label{sec:configuring_Nentry_RFMth}
There are multiple possible \name configurations for a single target $\nth$ because both $\entrynum$ and $\rfmth$ can change to satisfy $M < \nth/2$.
Figure~\ref{fig:4_nentry_rfmth} plots ($\entrynum$, $\rfmth$) pairs that satisfy this condition for various $\nth$ values (e.g., 1.5K, 3.125K, ..., 50K).
First, a trade-off is depicted between $\entrynum$ and $\rfmth$ regardless of $\nth$.
The decreased $\entrynum$ implies less area usage but results in a lower $\rfmth$, incurring more performance and energy overhead due to more frequent issuing of RFM commands.
This trade-off exists for all instances of $\nth$, but the appearance of the curve differs across various $\nth$ values.
A scheme similar to \name but based on a Lossy-counting algorithm is also noted at $\nth$ values of 50K and 25K, which clearly demonstrates a larger table for a given $\nth$.

When $\nth$ is sufficiently high (e.g., larger than 12.5K), it is possible to set $\rfmth$ to approximately 256 at a relatively small $\entrynum$.
Then, \name can achieve RH-protection with relatively low area, performance, and energy overhead.
In contrast, when $\nth$ is low, maintaining the low performance/energy overhead (i.e., sufficiently large $\rfmth$) requires a substantially larger $\entrynum$.
Overall, this is a trade-off that a DRAM vendor must consider when determining $\entrynum$.
The target $\nth$ level can be adjusted by tweaking the $\rfmth$ value even if $\entrynum$ is fixed.
This flexibility can be handy when the scheme must be built based on the predicted $\nth$ level and thus a fixed area, as it can avoid excessive performance/energy overhead.

\subsection{Wrapping Mithril Counters} 
\label{sec:hardware_implementation}
The absolute counter value of the \name table can increase in an unbounded manner during its run-time, which complicates the hardware implementation. 
Prior works solved this issue by periodically resetting the entire table~\cite{micro-2020-graphene, TWiCe} or by using a duplicate counter table in an interleaving fashion~\cite{blockhammer}; these two strategies lead to two-fold degradation of the predefined threshold level (from $\nth/2$ to $\nth/4$) and the area, respectively.
However, \name can avoid this.
Unlike prior approaches, \name does not require the absolute value of the estimated count.
Instead, we require the \emph{relative difference} of the estimated count in the minimum estimated count on the \name table.
Moreover, due to the operational behavior of \name, the maximum difference between the $\maxptr$ and $\minptr$ counter values is always bounded.
Therefore, we adopt a wrapping counter for \name table implementation.
If we provision enough bits capable of expressing a value larger than the maximum difference in the table, the wrapping counter can always correctly identify the relative size relationship among \name table entries.
Through this implementation, we acquire a two-fold benefit.

\setcounter{figure}{5}
\begin{figure}[!tb]
    \begin{center}
    \includegraphics[width=0.99\columnwidth]{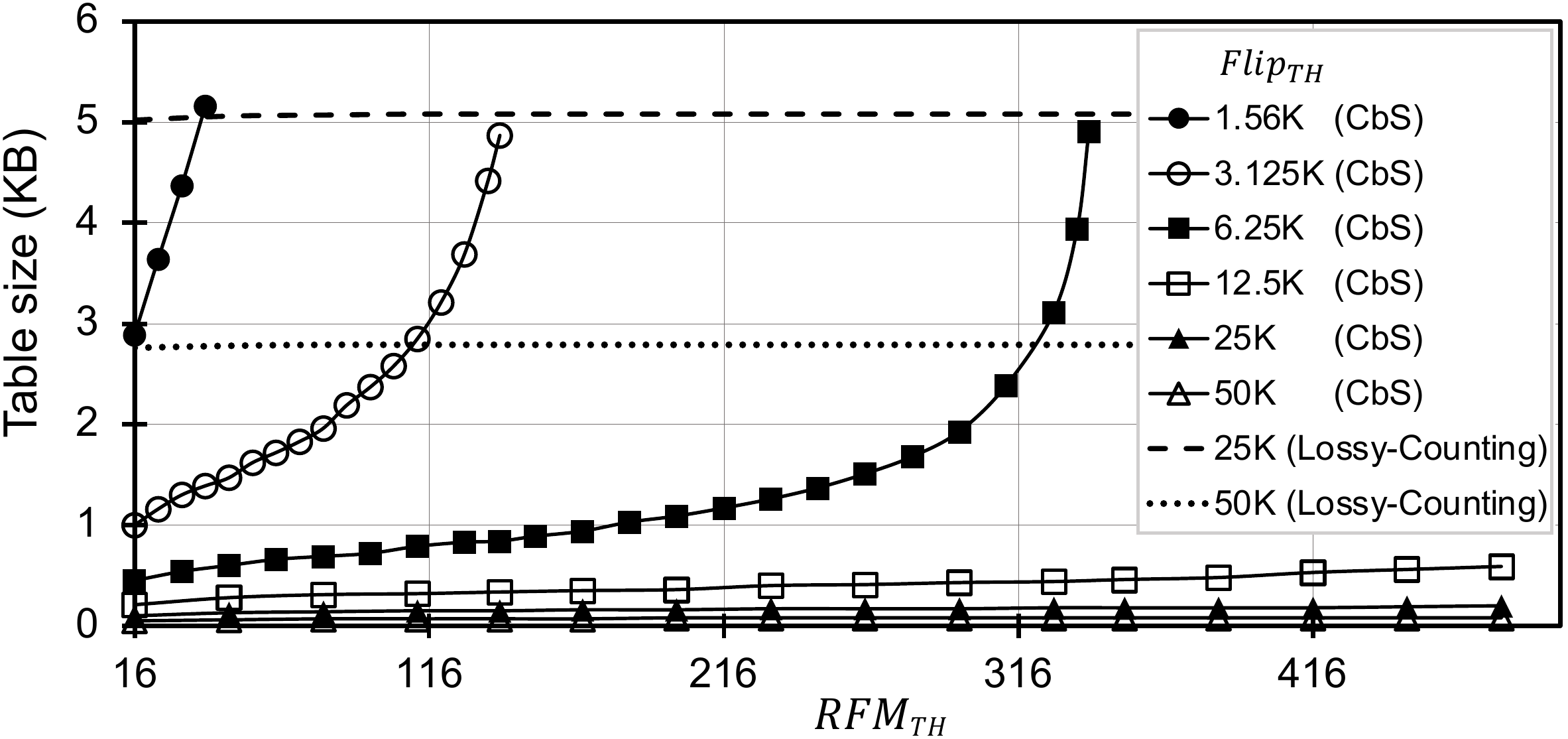}
    \end{center}
    \vspace{-0.2in}
    \caption{Each line denotes the possible configuration of $\entrynum$ (represented by the counter table size) and $\rfmth$ that can protect victim rows against RH at the given $\nth$ value. The RFM-based scheme built with a Lossy-Counting algorithm is also indicated by the dotted lines.}
    \label{fig:4_nentry_rfmth}
    \vspace{-0.1in}
\end{figure}

\section{Enhancing \name Further}
\label{sec:contribution-3}

\subsection{Adaptive Refresh}
\label{sec:adaptive}

Section~\ref{sec:contribution-2} assumed that \name performs a preventive refresh for every RFM command.
However, if Mithril can successfully distinguish a benign memory access pattern from an RH attack pattern, we can skip some of the RFM commands.
We find that the difference between the $\maxptr$ and the $\minptr$ count values is an effective identifier of such different patterns. Thus, we propose to perform a preventive refresh only when this difference exceeds a certain threshold ($\adth$).
This is referred to as an \emph{adaptive refresh policy}.

\begin{figure}[tb!]
    \begin{center}
    \includegraphics[width=1\columnwidth]{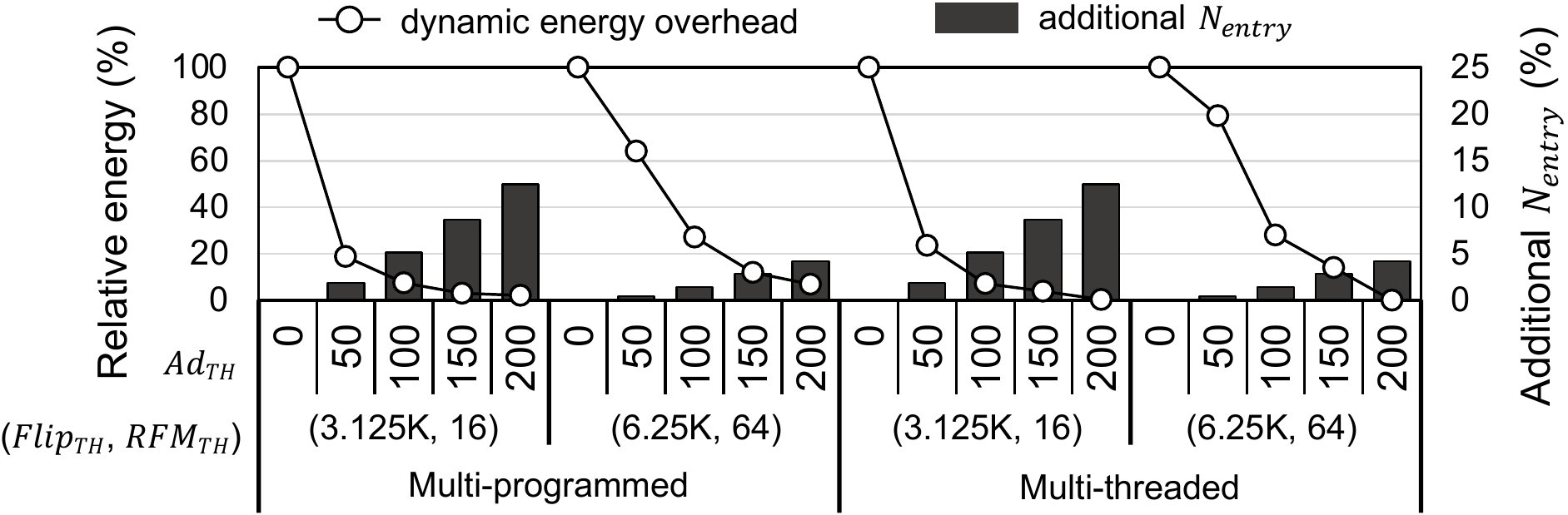}
    \end{center}
    \vspace{-0.15in}
    \caption{Relative dynamic energy overhead and additional $\entrynum$ against the default Mithril (0 $\adth$) for four different $\adth$ levels. 
    }
    \vspace{-0.05in}
    \label{fig:5_cont3_various_diffs}
\end{figure}

The difference between the $\maxptr$ and the $\minptr$ count values serves as a decent proxy of possible RH attacks, as large difference implies a high concentration of memory accesses to a small number of rows.
Therefore, if $\adth$ is set large enough, \name with the adaptive refresh policy can effectively filter out the ACT patterns observed by normal workloads.
Figure~\ref{fig:5_cont3_various_diffs} shows the effectiveness of the adaptive refresh policy, nearly eliminating additional energy overhead with benign workloads (see Section~\ref{sec:analysis} for the details of the experimental setup).

\begin{figure}[tb!]
    \begin{center}
    \includegraphics[width=0.9\columnwidth]{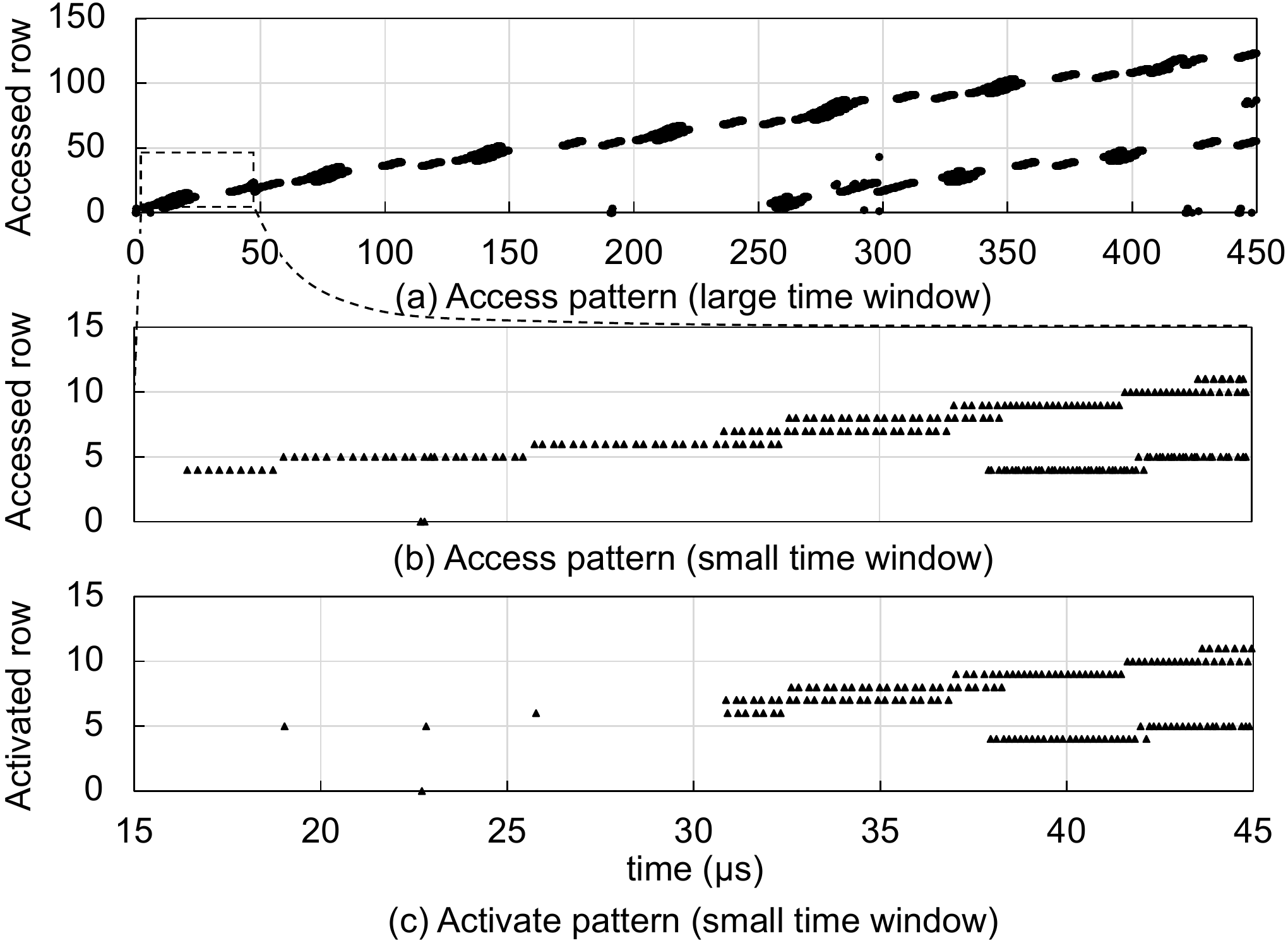}
    \end{center}
    \vspace{-0.15in}
    \caption{Example of a large object sweep pattern of lbm in SPEC CPU2017: (a) the memory access pattern in the large time window, (b) magnified to a small window, (c) the activation pattern in the small window.}
    \vspace{-0.12in}
    \label{fig:5_cont3_normal_workload.pdf}
\end{figure}

Among the multiple $\adth$ values, we can identify that the adaptive refresh policy is effective at the range of 100 to 200 in all cases.
We seek the root cause in the cross-play of memory access patterns of ordinary workloads and the DRAM row size.
Multithreaded or memory-intensive workloads often exhibit large-object-sweep behavior that results in main-memory accesses (Figure~\ref{fig:5_cont3_normal_workload.pdf}(a)).
In such a case, memory accesses are concentrated on a small number of rows in a short time period (Figure~\ref{fig:5_cont3_normal_workload.pdf}(b)) while being rather evenly distributed over the entire footprint overall. Although such an access pattern may possess high DRAM row locality, inter-process/thread conflicts can cause a high rate of ACT per memory access (Figure~\ref{fig:5_cont3_normal_workload.pdf}(c)).
Here, the number of concentrated ACTs would be similar to the number of streaming RDs/WRs, which would be 128 for an 8KB DRAM row and a 64B cache line size.
This matches the range of the effective adaptive threshold values, although the exact value must be determined empirically.

The adaptive refresh policy causes a slight deterioration of the bound $\maxcount$ (Theorem 1), thus inducing a higher area or performance cost to ensure the same effect as the baseline.
However, such an effect is minimal unless $\adth$ is very high. Figure~\ref{fig:5_cont3_various_diffs} shows a small increase in $\entrynum$, a maximum of 12\% at only a very low $\nth$ value.
Proof of the adjusted bound can be derived from Theorem 1 but is omitted here due to a lack of space.

\subsection{\nameplus}
\label{sec:mithril_plus}
The adaptive refresh policy allows \name to skip a preventive refresh even when the RFM command is issued by the memory controller.
By doing so, \name can reduce energy consumption but not the performance overhead.
Regardless of whether a DRAM component actually performs refreshes, the MC will continue to issue RFM commands at every $\rfmth$ ACT.

Inspired by such a limitation, we propose an optional, more invasive extension of \name, termed \nameplus, which prevents the MC from issuing unnecessary RFM commands. 
\nameplus utilizes the mode register in the DRAM device, which is flagged when the difference between $\maxptr$ and $\minptr$ is smaller than the values of $\adth$.
At every $\rfmth$, MC reads the flag using the JEDEC-standard MRR (Mode Register Read) command, determining whether or not to issue the RFM command.
With this interface, \nameplus can substantially minimize the performance overhead in the common case of ordinary workloads at the expense of a modification to the RFM interface.

\subsection{Non-adjacent Row Hammer}
\label{sec:different_tRFM}
\name can follow approaches similar to those in prior works~\cite{micro-2020-graphene,blockhammer} with regard to handling a non-adjacent RH by adjusting the $M$ value and the number of rows required to execute a preventive refresh.
When the range of the RH effect is one (double-sided attack, which we have assumed thus far for Mithril), $M$ smaller than $\nth/2$ is safe.
However, when the range is broader, $M$ must be smaller than $\nth/(aggregated\ RH\ effect)$ for non-adjacent aggressors. Within the range of 3, the aggregated RH effect is 3.5~\cite{blockhammer}, with six victim rows to execute a preventive refresh.

\section{Evaluation}
\label{sec:analysis}

\begin{table}[!tb]
\centering
\small
\caption{Architectural parameters for simulation}
{
\resizebox{0.48\textwidth}{!}{
    \begin{tabular}{P{2.5cm}P{4.9cm}}
      \Xhline{3\arrayrulewidth}
      \multicolumn{2}{c}{{\bf Core Configurations (16 cores)}} \\ \Xhline{1.5\arrayrulewidth}
      Core & 3.6 GHz 4-way OOO cores \\
      LLC & 16 MB \\
      \Xhline{1.5\arrayrulewidth}
     \multicolumn{2}{c}{{\bf Memory System Configurations}} \\ \Xhline{1.5\arrayrulewidth}
     Module & DDR5-4800 \\
     Channel & 2 channels \\
     Configuration & 1 rank; 32 banks per rank \\
     Scheduling & BLISS \cite{BLISS} \\
     Page-Policy & Minimalist-open \cite{Minimalist-open} \\ 
     \texttt{tRFC}, \texttt{tRC}, \texttt{tRFM} & 295 ns, 48.64 ns, 97.28 ns \\
     \texttt{tRCD}, \texttt{tRP}, \texttt{tCL} & 16.64 ns \\
      \Xhline{3\arrayrulewidth}
    \end{tabular}\vspace{0.5em}
}
}
\label{tab:6_experiment_setup}
\vspace{-0.1in}
\end{table}

We evaluate the performance, energy, and area overhead of \name and \nameplus in comparison with the RFM-interface-compatible PARFM and BlockHammer, as well as the RFM-interface-non-compatible PARA, CBT, TWiCe, and Graphene.

\subsection{Experimental Setup}
\label{sec:experimental_setup}

\noindent
\textbf{Methodology:}
The performance overhead is evaluated based on McSimA+~\cite{McSimA+}. 
Table~\ref{tab:6_experiment_setup} summarizes the experimental setup. 
We use the normalized aggregate IPC as the performance metric, where the baseline is the aggregate IPC without applying any RH-protection scheme for a workload.
We count the number of ACTs, PREs, and executed preventive refreshes to calculate the dynamic energy dissipation. 
First, we synthesize the RTL implementation of the \name module using the TSMC 40 nm standard cell library with the Synopsys Design Compiler.
The area overhead is scaled down to DRAM 20 nm and then again scaled up 10$\times$~\cite{upmem_dram_process} to conservatively take the inferior DRAM process into account.
The hardware energy consumption of \name is also derived from the synthesis.

\noindent
\textbf{Workloads:}
We use 1) normal, 2) multi-sided RH, and 3) BlockHammer-performance-adversarial workloads for evaluation.
We use both multi-programmed and multi-threaded workloads for normal workloads, reporting their geo-mean values.
From SPEC CPU2017, we extract 100M instruction traces~\cite{simpoint} and render two different workloads, mix-high and mix-blend, each of which comprises 16 traces of memory-intensive and randomly selected workloads, respectively.
We execute 400M instructions in total. We also evaluate three different multi-threaded benchmarks (FFT and RADIX from SPLASH-2~\cite{SPLASH-2} and PageRank from GAP~\cite{GAP_Benchmark}).

We configure a multi-sided RH attack that targets multiple victims~\cite{trrespass,half-double},  typically 32 in total.
The adversarial pattern for BlockHammer in performance is configured to blacklist specific profiled rows that share the CBF (counting bloom filter) entry with the benign threads.
Each is activated just enough to reach the blacklist threshold.
This effectively throttles benign workloads, especially memory-intensive types.
Each RH attack or adversarial pattern runs simultaneously with the 15 other benign workloads.

\noindent
\textbf{Configurations:}
We select up to three different \name and \nameplus ($\entrynum$, $\rfmth$) configurations for each $\nth$, ranging from 50K to 1.5K. Recently observed\mbox{~\cite{revisiting}} $\nth$ values are approximately 5K, but 1.5K is reachable considering the continued scaling of process technology and the non-adjacent RH. At high $\nth$ values of 50K and 25K, $\rfmth$ at fixed to 256 given that $\entrynum$ is already low. 
At the lowest $\nth$ of 1.5K, $\rfmth$ is fixed at 32 because a higher $\rfmth$ value results in an overly high $\entrynum$.
We use a value of 200 for $\adth$ as the default value.
For PARFM, $\rfmth$ is fixed to satisfy a failure probability of $10^{-15}$ (a typical consumer memory reliability target\mbox{~\cite{blockhammer,cai-2017-error_characterization_flash,cai-2012-error_MLC_NAND,JEDEC_failure_mechanism,luo-2016-flash_modeling,patel-2017-reach_profiler_DRAM_retention})} for 64 banks within a 32ms time period (tREFW) for each $\nth$.
The probability degrades if the number of banks to support increases.

We reconfigure BlockHammer\footnote{
BlockHammer uses a pair of interleaved counting bloom filters (CBFs) similar to Count-min Sketch algorithm. Each CBF is reset at every CBF lifetime ($t_{CBF}$), which typically matches tREFW. There exists a certain blacklist threshold ($N_{BL}$) of ACT that triggers a \emph{delay} on a certain \emph{row} when it is surpassed. The delay time ($t_{Delay}$) is calculated as $(t_{CBF}-N_{BL}{\times}\text{tRC})/(\nth-N_{BL})$. Thread-level scheduling support is built on top of these to throttle the aggressor \emph{thread} itself.
} 
to match our simulation environment and our target $\nth$ values. For (CBF size, $N_{BL}$) pairs, we used (1K, 17.1K), (1K, 8.6K), (1K, 4.3K), (2K, 2.1K), (4K, 1.1K), and (8K, 0.49K) for $\nth$ from 50K to 1.5K.
Under our system of four banks per thread, the number of ACTs per row easily exceeds 700 (as opposed to 109 ACTs in the original BlockHammer system with more banks per thread~\cite{blockhammer}), especially for memory-intensive workloads. 
Because $N_{BL}$ must be lower than $\nth/2$ (750 for a $\nth$ value of 1.5K), it is difficult to set an appropriate $N_{BL}$ value that distinguishes benign accesses from aggressor accesses and fulfill RH-protection at a $\nth$ value of 1.5K while also incurring minimal performance overhead.

Other prior schemes not compatible with RFM are also configured for a fair comparison with \name. TWiCe and Graphene are configured using the equations provided in each work to be applied to the DDR5 specification. PARA is configured to satisfy a failure probability of $10^{-15}$. CBT is configured to follow the configuration in the original work~\cite{isca-2017-cbt,ieee-cal-2017-cbt}.

\begin{figure}[tb!]
    \center
    \includegraphics[width=0.9\linewidth]{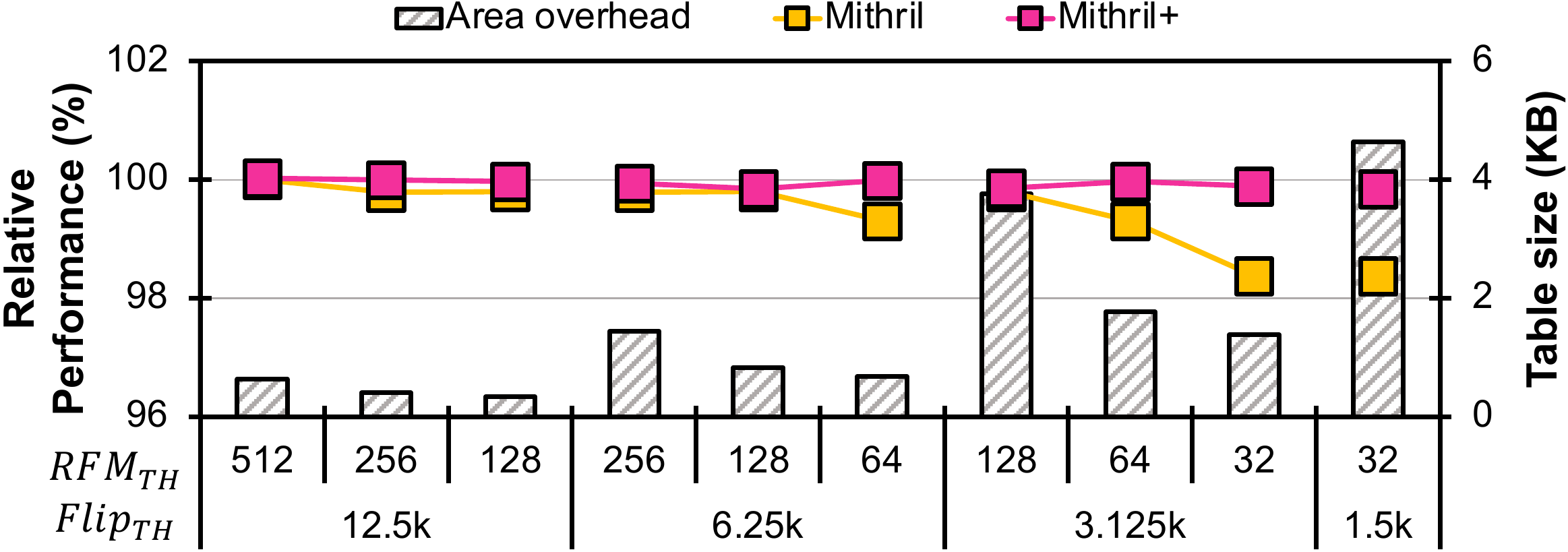}
    \vspace{-0.1in}
    \caption{The relative performance and area overhead of \name and \nameplus. 
    }
    \label{fig:6_1_mithril_config}
    \vspace{-0.12in}
\end{figure}

\subsection{The Overheads of \name and \nameplus}
\label{sec:rfm_based_schemes}
Mithril+ shows nearly zero performance overhead at all $\nth$ levels. The performance of Mithril degrades, with the amount depending on the target $\nth$ and $\rfmth$ configurations. 
There exists a performance-area trade-off for every $\nth$, which is amplified as $\nth$ value becomes smaller.

Mithril can support the recently observed $\nth$ values of approximately 6.25K\mbox{~\cite{revisiting}} with an $\rfmth$ of 128, which results in performance overhead of less than 0.5\% and a table size per bank of 1KB. Mithril can also support lower $\nth$ values, though at the cost of around 2\% of the performance and 4KB of area overhead. The area overhead of Mithril+ is identical to that of Mithril, with only negligible performance overhead.

\subsection{Comparison with Other Interface-Compatible Schemes}
\begin{figure}[tb!]
    \center
    \includegraphics[width=1.0\linewidth]{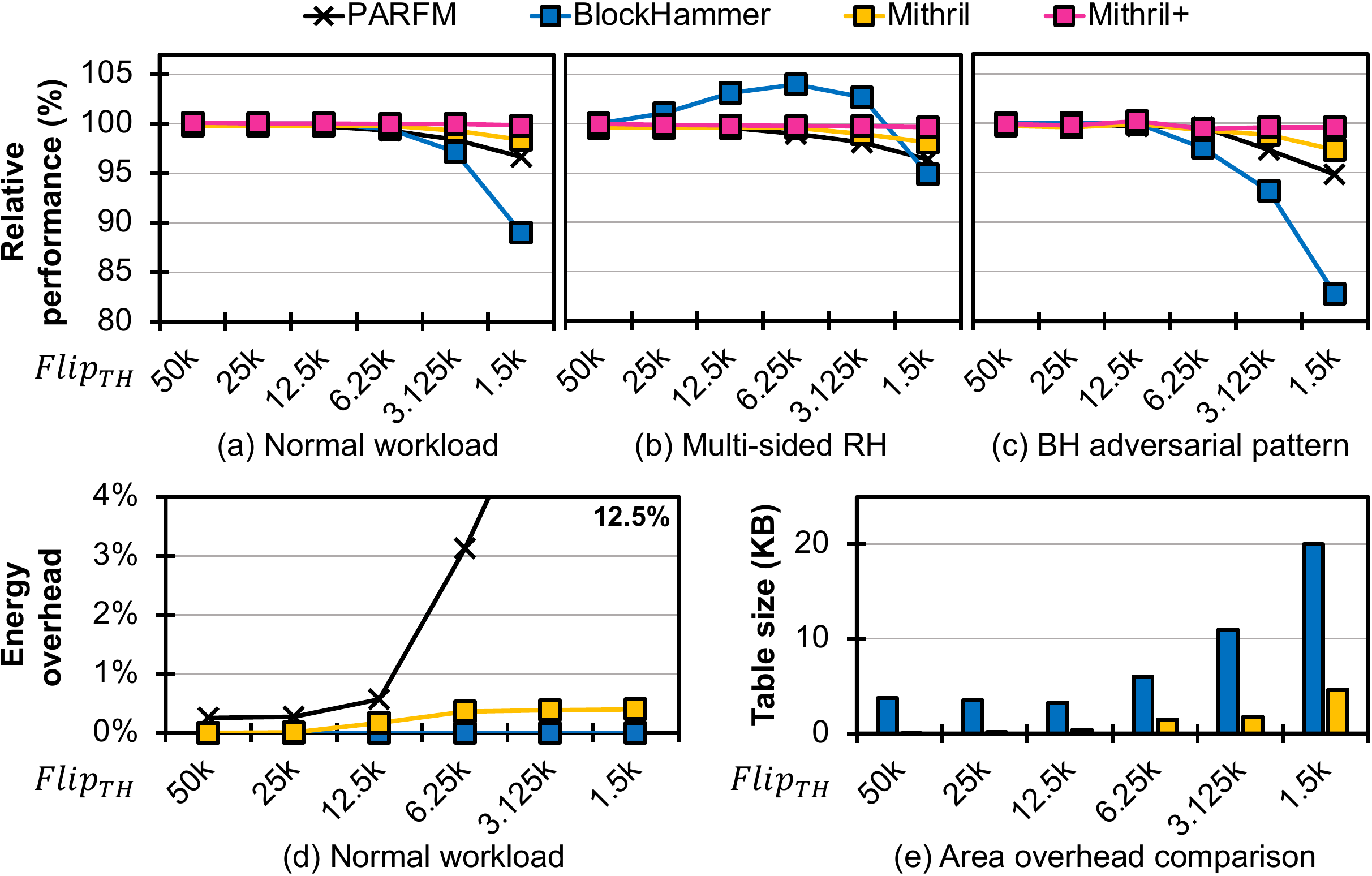}
    \vspace{-0.2in}
    \caption{(a), (b), and (c): Relative performance at normal workloads, under a multi-sided RH-attack, and BlockHammer-adversarial patterns; (d) Relative dynamic energy at normal workloads; and (e) Area overhead of BlockHammer and \name.}
    \label{fig:6_3_compatible}
    \vspace{-0.12in}
\end{figure}
\label{sec:6_comparison_compatible}
Figure~\ref{fig:6_3_compatible} shows the performance and the energy overhead of other RFM-interface-compatible schemes of PARFM and BlockHammer on multiple workloads for $\nth$ values ranging from 50K to 1.5K.
First, on normal workloads (Figure~\ref{fig:6_3_compatible}(a)), both Mithril+ and \emph{Mithril show small performance degradation of less than 2\%, superior to that of both PARFM and BlockHammer}. BlockHammer is particularly vulnerable at the low $\nth$ of 1.5K because it is prone to misidentifying benign threads and throttling them under such a condition.

Second, at the multi-sided RH (Figure~\ref{fig:6_3_compatible}(b)), BlockHammer exhibits a better aggregate IPC of up to 5\% for higher $\nth$ values, but it degrades again at a low $\nth$ value. This occurs because when BlockHammer successfully identifies RH attacking threads and throttles them, benign threads can benefit in return. However, this again leads to vulnerabilities during misclassifications when $\nth$ is lower than, for instance, 1.5K. The performance of \name and PARFM are agnostic with regard to the access patterns.

Lastly, regarding the performance of BlockHammer with an adversarial pattern (Figure~\ref{fig:6_3_compatible}(c)), the performance of BlockHammer degrades severely, with as much as a 17\% drop in the aggregate IPC. This implies the possibility of a critical \emph{performance (not RH) attack on systems equipped with BlockHammer, as its throttling feature works as a double-edged sword depending on how effectively it identifies RH attacking threads}.

\emph{The energy overhead of Mithril and Mithril+ are less than 0.4\%}, even when $\nth$ is 1.5K. These values are much smaller than that of PARFM and slightly higher than that of BlockHammer (Figure~\ref{fig:6_3_compatible}(d)). This occurs because the adaptive refresh policy successfully identifies ordinary workloads, skipping many of the RFM commands and not triggering additional preventive refreshes. PARFM shows the energy overhead in cases when every RFM command triggers a preventive refresh. BlockHammer causes only minimal logic energy because it is a throttling-based scheme.

\emph{The table size overhead of Mithril is much smaller than that of BlockHammer at all $\nth$} levels.
Figure~\ref{fig:6_3_compatible}(e) shows the table size overhead for each scheme. PARFM is omitted due to its negligible overhead, and that of Mithril+ is identical to Mithril. The table size of \name is up to 60$\times$ and a minimum of 4$\times$ smaller than that of BlockHammer at all $\nth$ levels. The table size comparison is discussed further in Section~\ref{sec:6_table_size_overhead}.

\subsection{Comparison with Interface Non-Compatible Schemes}
\begin{figure}[tb!]
    \center
    \includegraphics[width=1\linewidth]{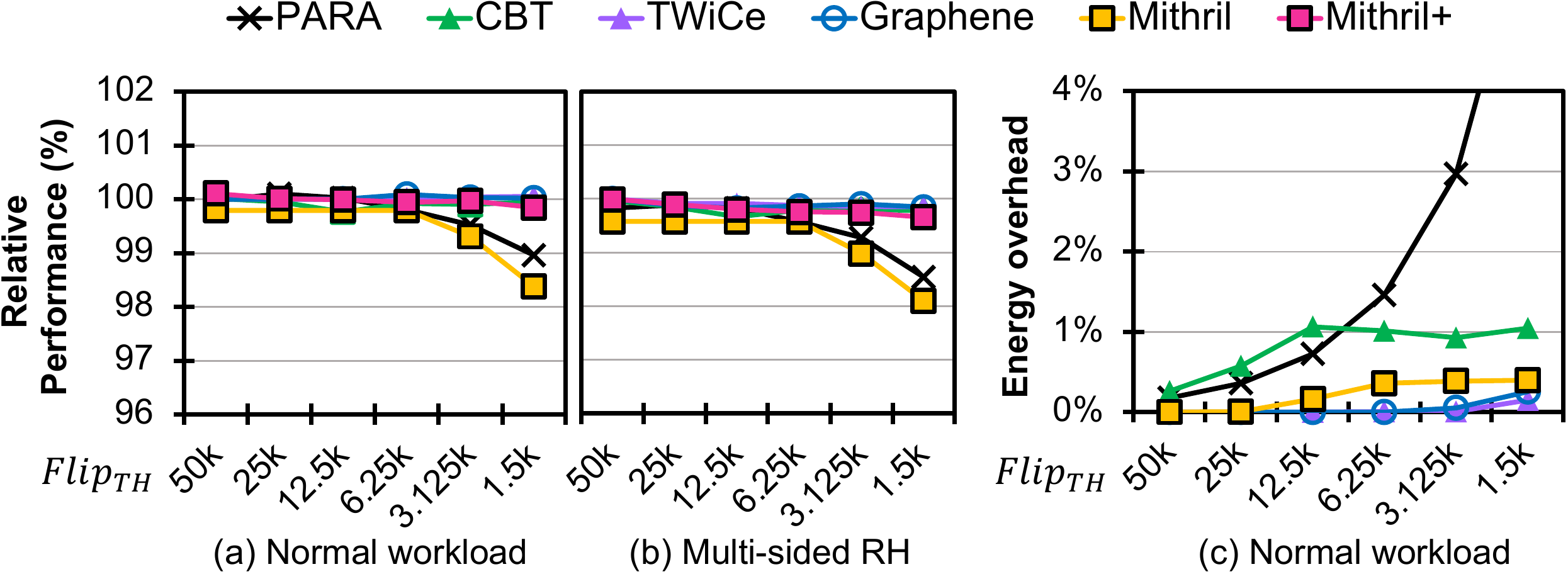}
    \vspace{-0.25in}
    \caption{Relative performance at (a) normal workloads and (b) multi-sided RH-attack; (c): Relative dynamic energy at normal workloads.}
    \label{fig:6_2_not_compatible}
    \vspace{-0.1in}
\end{figure}
\label{sec:6_comparison_incompatible}
Mithril and Mithril+ also show \emph{competitive performance and energy overhead compared to the RFM-non-compatible prior works} of PARA, CBT, TWiCe, and Graphene. Under both normal workloads and a multi-sided RH attack situation (Figure~\ref{fig:6_2_not_compatible}(a), (b)), Mithril+ shows performance degradation of less than 0.2\%, comparable to those of TWiCe, Graphene, or CBT. The performance degradation of Mithril is worse than those of other schemes but is limited to less than 2\% even at the low $\nth$ of 1.5K. The energy overhead of Mithril is comparable to those of TWiCe and Graphene at less than 1\% even when $\nth$ is 1.5K (Figure~\ref{fig:6_2_not_compatible}(c)).

\subsection{Table Size Overhead}
\begin{table}[tb!]
\centering
\caption{Per Bank Table Size Comparison (KB)}
\label{tab:6_area_KB}
\resizebox{0.48\textwidth}{!}{
\begin{tabular}{l|rrrrrr} 
\Xhline{3\arrayrulewidth}
      \textbf{\begin{tabular}[c]{@{}l@{}}Scheme\end{tabular}}
    & \begin{tabular}[c]{@{}c@{}}50K\end{tabular}
    & \begin{tabular}[c]{@{}c@{}}25K\end{tabular}
    & \begin{tabular}[c]{@{}c@{}}12.5K\end{tabular}
    & \begin{tabular}[c]{@{}c@{}}6.25K\end{tabular}
    & \begin{tabular}[c]{@{}c@{}}3.125K\end{tabular}
    & \begin{tabular}[c]{@{}c@{}}1.5K\end{tabular} \\ 
\hline
 CBT @ MC              & 0.47 & 0.97 & 2.0\enspace     & 4.12  & 8.5\enspace    & 17.5\enspace  \\
                       Graphene @ MC     & 0.14 & 0.21 & 0.51  & 0.99  & 1.92   & 3.7\enspace   \\
                       BlockHammer @ MC     & 3.75 & 3.5\enspace  & 3.25  & 6.0\enspace     & 11.0\enspace     & 20.0\enspace    \\ 
 TWiCe @ buffer chip           & 2.79 & 5.08 & 9.54  & 18.27 & 35.29  & 71.26 \\ 
 \name-256 @ DRAM        & 0.08 & 0.17 & 0.41  & 1.45  & -\enspace\enspace      & -\enspace\enspace     \\
                       \name-128 @ DRAM        & 0.07 & 0.15 & 0.34  & 0.84  & 3.76   & -\enspace\enspace     \\
                       \name-64 @ DRAM         & 0.07 & 0.14 & 0.3\enspace   & 0.68  & 1.78   & -\enspace\enspace     \\
                       \name-32 @ DRAM        & 0.06 & 0.13 & 0.27  & 0.57  & 1.38   & 4.64  \\
\Xhline{3\arrayrulewidth}
\end{tabular}
}
\vspace{0.03in}
\\ * \name-(256/128/64/32) denote different $\rfmth$ values ranging from 256 to 32.
\vspace{-0.15in}
\end{table}

\label{sec:6_table_size_overhead}

We report the counter table size of each scheme in units of \emph{KB per bank} (see Table~\ref{tab:6_area_KB}).
While MC-side schemes benefit from their use of faster transistors, abundant wiring resources, and a relaxed area budget, the number of total banks is much higher (1,024), and the target $\nth$ must be pessimistic.
DRAM-side schemes benefit from fewer banks (32) to support per device and more accurate $\nth$ values, but they are hindered by slower transistors and a tighter area/wiring budget.

\name shows lower or competitive area overhead in terms of the KB per bank, reaching 0.024$mm^2$ when $\nth$ equals 6.25K.
This represents 1\% of a single DDR5 chip~\cite{ddr5-chipsize} when multiplied by 32 to cover 32 banks per chip. While both Graphene and Mithril share fundamentally the same CbS algorithm as their tracking mechanism, their table size overhead differs for several reasons. First, as an advantage for Mithril, it does not require a table reset due to the wrapping counter scheme, resulting in two-fold reduction. Also, the per-entry bit width of the counter CAM is smaller in Mithril because the maximum value is bounded to $M$ (Theorem 1), which is smaller than the Graphene case for maximum number of ACTs in the tREFW window. At a $\nth$ value of 1.5K, we ensure that $\rfmth$ is small to minimize the performance drop, resulting in increased $\entrynum$ and area overhead.

\section{Related Work}
\label{sec:related-work}

\noindent \textbf{\rh (RH) on Real Systems:} 
RH has been shown to be able to bypass all system memory protection schemes, allowing adversaries to compromise the confidentiality and integrity of actual systems. In 2015, Google~\cite{Google_Project_Zero} demonstrated that a user-level program could breach the system-level security of a typical PC by exploiting the RH vulnerability of the system. A number of successful attacks followed \cite{Google_Project_Zero}, including those compromising mobile devices~\cite{Drammer, mobile_attack1} and servers~\cite{RowHammer_JS, server_attack1, server_attack2}, thus breaking the authentication process and damaging the entire system, even when a system protects memory locations near sensitive data~\cite{micro-2020-pthammer}. Because RH undermines the fundamental principle of memory isolation, it has been regarded as a serious threat, drawing mitigation proposals from software, architecture, and hardware levels.

\noindent \textbf{Architectural Proposals to Mitigate RH:} There have been deterministic~\cite{CAT, cat-two, TWiCe, micro-2020-graphene,blockhammer} and probabilistic~\cite{PARA,PRoHIT, MRLoc} schemes proposed to mitigate RH attacks at the architecture level. Among these, \cite{MRLoc,PRoHIT,CAT} are susceptible to adversarial DRAM access patterns. TWiCe~\cite{TWiCe} and CAT-TWO~\cite{cat-two} are relatively free from this susceptibility but require an order of magnitude more storage to track aggressor rows compared to Graphene~\cite{micro-2020-graphene}. PARA~\cite{PARA} incurs low performance and energy overhead, whereas it is also extremely area-efficient as it does not require counters to trace aggressor rows. Yet, the protection is probabilistic in nature; even if the probability is quite small, there is a non-zero probability that a victim row will not be refreshed after reaching its RH threshold. BlockHammer~\cite{blockhammer} uses a throttling approach backed up with thread-level MC scheduling.

\section{Conclusion}
\label{sec:conclusion}

Here, we propose \name, a DRAM-side, RFM-compatible, efficient scheme that provides deterministic safety against \rh attacks.
First, we show that the conventional algorithms and methodologies used in previous architectural RH-prevention schemes are not compatible with the RFM command introduced in the latest DRAM specifications, such as DDR5 and LPDDR5.
By mathematical defining the maximum bound of activation count without a refresh in a tREFW time window, we guarantee safety at a specific $\nth$ value. 
The devised adaptive refresh policy decreases the energy overhead by exploiting the row activation patterns of ordinary workloads.
Moreover, we proposed \nameplus, which requires a slight modification of the RFM interface.
It utilizes the existing DRAM command to skip the sending of RFM commands, which can significantly reduce the performance overhead of \name. 
Our evaluation demonstrates that \name achieves a significantly low energy overhead in all cases compared to PARFM, whereas it incurs slightly higher performance overhead. 
\nameplus shows not only low energy overhead but also significantly lower performance overhead such that it is comparable to Graphene, a state-of-the-art RH-prevention scheme that does not support RFM.

\section*{Acknowledgment}
This work was supported by Institute of Information \& communications Technology Planning \& Evaluation (IITP) grant funded by the Korea government (MSIT) (2020-0-01300, Development of AI-specific Parallel High-speed Memory Interface, and 2021-0-01343, Artificial Intelligence Graduate School Program (Seoul National University)).
Jung Ho Ahn is the corresponding author.

\section{Appendix}
\label{sec:appendix}

\subsection{Proof for Theorem 1}
\label{sec:mithril_mathematical_proofs}

\noindent\textbf{Theorem 1.} \emph{Within any tREFW, an increase in the estimated count for any single row is bounded to $\maxcount$, which is a function of $\entrynum$ and $\rfmth$.}\\
\vspace{-0.1in}

\Scale[0.8]{
\begin{minipage}{1.0\columnwidth}
\begin{align*}
M = \sum_{k=1}^{\entrynum} \frac{\rfmth}{k} + \frac{\rfmth}{\entrynum} \left(\frac{\text{tREFW}(1-\frac{\text{tRFC}}{\text{tREFI}})}{\text{tRC} \times \rfmth+\text{tRFM}} - 2\right)
\end{align*}
\end{minipage}}\vspace{0.075in}\\

\begin{figure}[tb!]
    \begin{center}
    \includegraphics[width=1\columnwidth]{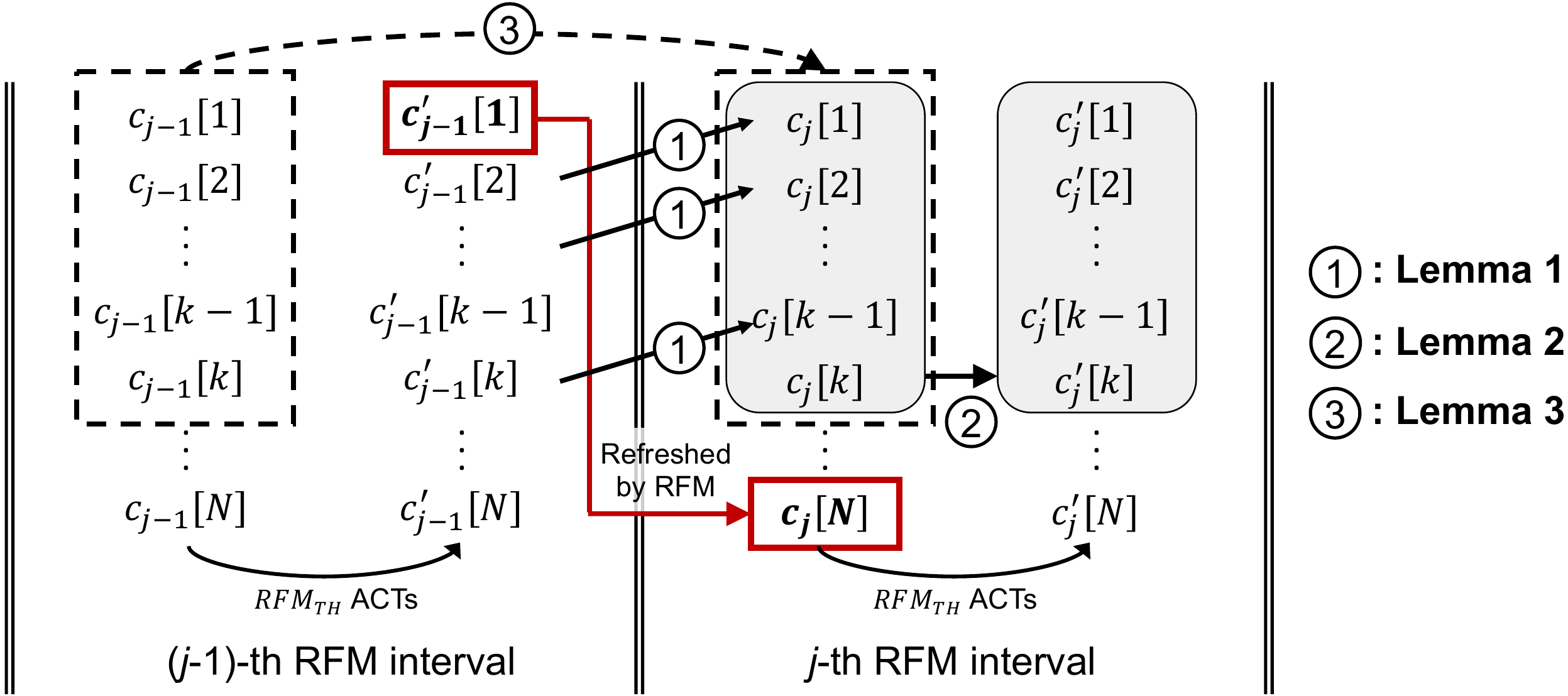}
    \end{center}
    \vspace{-0.15in}
    \caption{Example case illustrating how the estimated count is updated between two consecutive RFM intervals.
    }
    \vspace{-0.1in}
    \label{fig:notations_for_proof}
\end{figure}

Henceforth, $\entrynum$ is replaced by $N$. Also, $W$ represents the maximum number of RFM intervals (the period between two consecutive RFM commands) within a tREFW. It is computed as follows:
\vspace{-0.15in}

\Scale[0.75]{
\begin{minipage}{1.0\columnwidth}
\vspace{9pt}
\begin{align*}
W = \left\lceil ( \text{tREFW} - (\text{tREFW}/\text{tREFI}) \times \text{tRFC} ) / ( \text{tRC}\times\rfmth + \text{tRFM}) \right\rceil
\end{align*}
\end{minipage}}
\vspace{0.01in}

\noindent Suppose $c_{j}[i]$ is the $i$-th largest estimated count in the \name table at the beginning of the $j$-th RFM interval ($1 \leq i \leq N, 1 \leq j \leq W$).
$c'_{j}[i]$ is the $i$-th largest estimated count in the table at the end of the $j$-th RFM interval. 
Figure~\ref{fig:notations_for_proof} illustrates a prime example of such notations. Then, the following Lemmas hold true:

\vspace{0.075in}
\noindent\textbf{Lemma 1.} $c'_{j}[i] = c_{j+1}[i-1]$ \vspace{0.05in}\\
\textbf{Proof:} At the end of each RFM interval, one of the entries with the largest estimated count (i.e., $c'_{j}[1]$) becomes the target for the RFM refresh, and its estimated count is reset to the minimum count in the table.
Thus, the ranks of all other entries are increased by one after the RFM refresh. 

\vspace{0.1in}\noindent\textbf{Lemma 2.} $\sum_{i=1}^{k} c'_{j}[i] \leq \sum_{i=1}^{k} c_{j}[i] + \rfmth$ for \linebreak $1 \leq k \leq N $\vspace{0.05in}\\  
\textbf{Proof:} Considering that there are $\rfmth$ ACTs within each RFM interval and $c'_{j}[i]$ is larger than or equal to $c_{j}[i]$ for all values of $i$ by definition, the following holds true:
\vspace{-0.15in}\\
\Scale[0.74]{
\begin{minipage}{1.0\columnwidth}
\vspace{9pt}
\begin{align*}
\sum_{i=1}^{k} c'_{j}[i] &= \sum_{i=1}^{N} c'_{j}[i] - \sum_{i=k+1}^{N} c'_{j}[i]\\
&= \sum_{i=1}^{N} c_{j}[i] + \rfmth - \sum_{i=k+1}^{N} c'_{j}[i]\\
&= \sum_{i=1}^{k} c_{j}[i] - \sum_{i=k+1}^{N} (c'_{j}[i] - c_{j}[i]) + \rfmth\\
&\leq \sum_{i=1}^{k} c_{j}[i] + \rfmth
\end{align*}
\end{minipage}}

\vspace{0.1in}\noindent\textbf{Lemma 3.}  $\sum_{i=1}^{k} c_{j}[i] \leq \sum_{i=1}^{k} c_{j-1}[i] + \rfmth$ for \linebreak $1 \leq k \leq N $\vspace{0.05in}\\
\textbf{Proof:} This is an obvious extension of Lemma 2 because $\sum_{i=1}^{k} c_{j}[i] \leq \sum_{i=1}^{k} c'_{j-1}[i]$. In other words, an RFM refresh always decreases the sum of the top $k$ counter values in the table.

\vspace{0.1in}\noindent\textbf{Lemma 4.} $\sum_{i=1}^{k} c_{j}[i] \leq \frac{k}{k+1} (\sum_{i=1}^{k+1} c_{j-1}[i] + \rfmth)$ \vspace{0.05in} for $1 \leq k \leq N-1$\\
\textbf{Proof:} Using Lemma 1, Lemma 2, and the fact that $c'_{j-1}[1] \geq c'_{j-1}[i]$ for all $i$, the following holds true:
\vspace{-0.09in}\\
\Scale[0.8]{
\begin{minipage}{1.0\columnwidth}
\vspace{9pt}
\begin{align*}
\sum_{i=1}^{k} c_{j}[i] &= \sum_{i=2}^{k+1} c'_{j-1}[i]\\
&= \sum_{i=1}^{k+1} c'_{j-1}[i] - c'_{j-1}[1]\\
&\leq \sum_{i=1}^{k+1} c'_{j-1}[i] - \frac{1}{k+1} \sum_{i=1}^{k+1} c'_{j-1}[i]\\
&=\; \frac{k}{k+1} \sum_{i=1}^{k+1} c'_{j-1}[i]\\
&\leq\; \frac{k}{k+1} \left(\sum_{i=1}^{k+1} c_{j-1}[i] + \rfmth\right)\\
\end{align*}
\vspace{-0.3in}
\end{minipage}}
    
With these Lemmas, we are ready to prove Theorem 1. Proving Theorem 1 is equivalent to proving the following: \vspace{-0.1in}\\
\Scale[0.93]{
\begin{minipage}{1.0\columnwidth}
\vspace{9pt}
\begin{align*}
c'_{W}[1] - c_{1}[N] \leq M \text{~for given~} c_{1}[1], ... ,  c_{1}[N]
\end{align*}
\end{minipage}}
\vspace{0.08in}

\noindent This works because any row's estimated count that is increased during the $W$ RFM intervals is obviously less than the difference between the largest estimated count at the end ($c'_{W}[1]$) and the smallest estimated count at the beginning ($c_{1}[N]$). Accordingly, we can obtain the upper bound for $c'_{W}[1]$ as follows: 

\Scale[0.78]{
\begin{minipage}{1.0\columnwidth}
\vspace{-0.08in}
\begin{align*}
c'_{W}[1] &\leq c_{W}[1] + \rfmth \;\;\; (\because \text{~Lemma 2}) \\
&\leq \frac{1}{2} \left(\sum_{i=1}^2 c_{W-1}[i] + \rfmth\right) + \rfmth \;\;\; (\because \text{~Lemma 4}) \\
&\leq \frac{1}{3} \left(\sum_{i=1}^{3} c_{W-2}[i] + \rfmth\right) + \sum_{k=1}^{2} \frac{\rfmth}{k} \;\;\; (\because \text{~Lemma 4}) \\
\end{align*}
\vspace{-0.3in}
\end{minipage}}

\noindent Repeatedly applying Lemma 4 for a total of $N-1$ times, we obtain the following inequality:

\Scale[0.825]{
\begin{minipage}{1.0\columnwidth}
\vspace{-3pt}
\begin{align*}
c'_{W}[1] &\leq \frac{1}{N} \left(\sum_{i=1}^{N} c_{W-N+1}[i] + \rfmth\right) + \sum_{k=1}^{N - 1} \frac{\rfmth}{k}\\
&= \frac{1}{N} \left(\sum_{i=1}^{N} c_{W-N+1}[i]\right) + \sum_{k=1}^{N} \frac{\rfmth}{k}\\
\end{align*}
\vspace{-0.3in}
\end{minipage}}

\noindent At this point, we can no longer apply Lemma 4 and instead apply Lemma 3 ($k = N$) $W-N$ times. 

\Scale[0.825]{
\begin{minipage}{1.0\columnwidth}
\vspace{-3pt}
\begin{align*}
c'_{W}[1] &\leq \frac{\sum_{i=1}^{N} c_{1}[i]}{N} + \frac{(W-N)\rfmth}{N} + \sum_{k=1}^{N}\frac{\rfmth}{k}\\
& = \frac{\sum_{i=1}^{N} c_{1}[i]}{N} + M - \frac{N-2}{N}\rfmth
\end{align*}
\vspace{-0.3in}
\end{minipage}}\vspace{0.1in}\\

\noindent Earlier, we showed that proving Theorem 1 is equivalent to proving $c'_{W}[1] - c_{1}[N] \leq M$. With the above equation, proving the following is the only step left to prove Theorem~1:

\Scale[0.9]{
\begin{minipage}{1.0\columnwidth}
\vspace{-1.45in}
\begin{align*}
\frac{\sum_{i=1}^{N} c_{1}[i]}{N} - c_{1}[N] \leq \frac{N-2}{N} \rfmth 
\end{align*}
\vspace{-1.8in}
\end{minipage}}\vspace{0.1in}\\

\noindent Here, the left-hand side can be represented as follows. 

\Scale[0.8]{
\begin{minipage}{1.0\columnwidth}
\vspace{0pt}
\begin{align*}
\frac{\sum_{i=1}^{N} c_{1}[i]}{N} -  c_{1}[N] =
\frac{\sum_{i=1}^{N} c_{1}[i] - N c_{1}[N]}{N} =
\frac{\sum_{i=1}^{N} (c_{1}[i] - c_{1}[N])}{N} \\
\end{align*}
\vspace{-0.2in}
\end{minipage}}

\noindent The upper bound of \begin{footnotesize}$\sum_{i=1}^{N} (c_{j}[i] - c_{j}[N])$\end{footnotesize} for any $j$-th RFM interval can be obtained by contradiction.
We assume that \begin{footnotesize}$\sum_{i=1}^{N} (c_{j}[i] - c_{j}[N])$\end{footnotesize} is maximized when $j$ is $m$ and that the difference between $c_{m}[1]$ and $c_{m}[N]$ is greater than $\rfmth$. Then, the following holds:

\Scale[0.9]{
\begin{minipage}{1.0\columnwidth}
\vspace{0pt}
\setcounter{equation}{2}
\begin{align} \label{eq:diff_c1_cN}
\begin{split}
c'_{m-1}[1] - c'_{m-1}[N]
&\;\geq\; c'_{m-1}[2] - c'_{m-1}[N] \\
&\;=\; c_{m}[1] - c_{m}[N]
\;>\; \rfmth \\
\end{split}
\end{align}
\vspace{-3pt}
\end{minipage}}

\noindent At the end of the ($m$-1)-th RFM interval, $c'_{m-1}[1]$ is reduced to $c'_{m-1}[N]$ by RFM. Therefore

\Scale[0.68]{
\begin{minipage}{1.0\columnwidth}
\vspace{3pt}
\begin{align*}
\sum_{i=1}^{N} (c_{m}[i] - c_{m}[N])
&= \sum_{i=1}^{N} (c'_{m-1}[i] - c'_{m-1}[N]) - (c'_{m-1}[1] - c'_{m-1}[N]) \\
&= \sum_{i=1}^{N} (c_{m-1}[i] - c_{m-1}[N]) + \rfmth - (c'_{m-1}[1] - c'_{m-1}[N]) \\
&< \sum_{i=1}^{N} (c_{m-1}[i] - c_{m-1}[N]) \;\;\; (\because ~\text{(\ref{eq:diff_c1_cN})})
\end{align*}
\vspace{3pt}
\end{minipage}}

\noindent This contradicts the contention that \begin{footnotesize}$\sum_{i=1}^{N} (c_{j}[i] - c_{j}[N])$\end{footnotesize} is maximized when $j$ is $m$. Therefore, if \begin{footnotesize}$\sum_{i=1}^{N} (c_{j}[i] - c_{j}[N])$\end{footnotesize} is maximized when $j$ is $m$, the difference between $c_{m}[1]$ and $c_{m}[N]$ is less than or equal to $\rfmth$. Then, we obtain the following inequality:

\Scale[0.75]{
\begin{minipage}{1.0\columnwidth}
\begin{align*}
\frac{\sum_{i=1}^{N} (c_{1}[i] - c_{1}[N])}{N} 
&\leq \frac{\sum_{i=1}^{N} (c_{m}[i] - c_{m}[N])}{N}\\
&= \frac{\sum_{i=1}^{N-2} (c_{m}[i] - c_{m}[N])}{N} \;\;\;  (\because ~c_{m}[N-1] = c_{m}[N])\\
&\leq \frac{\sum_{i=1}^{N-2} (c_{m}[1] - c_{m}[N])}{N}\\
&\leq \frac{(N-2)\rfmth}{N}
\end{align*}
\end{minipage}}

\subsection{Finding New $M$ for Adaptive Refresh}
\label{sec:adaptive_appendix}

If the adaptive refresh policy (Section~\ref{sec:adaptive}) is applied to \name, Lemmas 1 and 4 in Section~\ref{sec:mithril_mathematical_proofs} no longer hold because the preventive refresh may not occur at the end of the RFM interval.
The modified $M$ (henceforth $M'$) for the adaptive refresh is as follows.\\

\noindent\textbf{Theorem 2.} \emph{When the adaptive refresh is applied to \name, an increase in the estimated count for any single row within any tREFW is bounded to $M'$, which is a function of $\entrynum$, $\rfmth$, and $\adth$.}

\Scale[0.7]{
\begin{minipage}{1.0\columnwidth}
\begin{align*}
\maxcount' &= \sum_{k=1}^{n} \frac{\rfmth}{k} + \frac{(W-n^*+\entrynum-2) \rfmth + (\entrynum - n^*) \adth}{\entrynum}\\
&*\; W = \left\lceil ( \text{tREFW} - (\text{tREFW}/\text{tREFI}) \times \text{tRFC} ) / ( \text{tRC}\times\rfmth + \text{tRFM}) \right\rceil \\
&*\; n^* = \lceil (\entrynum \times \rfmth) / (\rfmth+\adth) \rceil
\end{align*}
\end{minipage}}\vspace{0.075in}\\

At the end of any $j$-th RFM interval, the preventive refresh does not occur if the difference between $c'_{j}[1]$ and $c'_{j}[N]$ is less than $\adth$.
Considering that the preventive refresh may not occur, we modify Lemma 4 to Lemmas 5 and 6 as follows: \\

\vspace{-0.1in}
\noindent\textbf{Lemma 5.} If $c'_{j-1}[1] - c'_{j-1}[N] > \adth$, then \\
$\sum_{i=1}^{k} c_{j}[i] \leq \frac{k}{k+1} (\sum_{i=1}^{k+1} c_{j-1}[i] + \rfmth)$ for \linebreak $1 \leq k \leq N-1$

\noindent\textbf{Proof:} If $c'_{j-1}[1] - c'_{j-1}[N] > \adth$, the preventive refresh occurs at the end of the ($j$-1)-th RFM interval, so we can derive the same result as Lemma 4.\\

\vspace{-0.1in}
\noindent\textbf{Lemma 6.} If $c'_{j-1}[1] - c'_{j-1}[N] \leq \adth$, then \\
\Scale[0.95]{
\begin{minipage}{1.0\columnwidth}
$\sum_{i=1}^{k} c_{j}[i] \leq \frac{k}{N} \left(\sum_{i=1}^{N} c_{j-1}[i] + \rfmth  + (N - k) \adth\right)$ for $1 \leq k \leq N-1$
\end{minipage}}

\noindent\textbf{Proof:} Because RFM does not occur at the $j$-th RFM interval, $c_{j}[i] = c'_{j-1}[i]$ for $1 \leq i \leq N$. Then, the following holds true.

\Scale[0.9]{
\begin{minipage}{1.0\columnwidth}
\begin{align*}
\sum_{i=1}^{N} c_{j}[i]
&= \sum_{i=1}^{N} c'_{j-1}[i]\\
&= \sum_{i=1}^{k} c'_{j-1}[i] + \sum_{i=k+1}^{N} c'_{j-1}[i]\\
&\geq \sum_{i=1}^{k} c'_{j-1}[i] + (N-k) c'_{j-1}[N]\\
&\geq \sum_{i=1}^{k} c'_{j-1}[i] + (N-k) (c'_{j-1}[1] - \adth)\\
&\geq \sum_{i=1}^{k} c'_{j-1}[i] + (N-k) \left(\frac{1}{k} \sum_{i=1}^{k} c'_{j-1}[i] - \adth\right)\\
&\geq \frac{N}{k} \sum_{i=1}^{k} c'_{j-1}[i] - (N-k) \adth\\
\sum_{i=1}^{k} c_{j}[i]
&= \sum_{i=1}^{k} c'_{j-1}[i]\\
&\leq \frac{k}{N} \left(\sum_{i=1}^{N} c'_{j-1}[i] + (N-k) \adth\right) \\
&\leq \frac{k}{N} \left(\sum_{i=1}^{N} c_{j-1}[i] + \rfmth + (N-k) \adth\right)\\
\end{align*}
\end{minipage}}

Similar to Theorem 1, proving Theorem 2 is equivalent to proving $c'_{W}[1] - c_{1}[N] \leq M'$.
For some arbitrary number $n$ smaller than $W$, assume that the preventive refresh does not occur at the $n$-th last RFM interval (i.e., the ($W-n$)\mbox{-}th RFM interval) and that the preventive refresh occurs at all the subsequent RFM intervals.
Even with the adaptive refresh, Lemmas 2 and 3 are still true.

If $n$ is greater than $N$, the upper bound of $c'_{W}[1] - c_{1}[N]$ is equivalent to $M$ (the result of Theorem 1). 
We can obtain this result by applying Lemma 5 (equivalent to Lemma 4) for $N-1$ times and applying Lemma 3 for $W-N$ times to $c'_{W}[1]$.

Otherwise, if $n$ is less than or equal to $N$, we first repeatedly apply Lemma 5 for a total of $n-1$ times to obtain the upper bound of $c'_{W}[1]$: \\

\vspace{-0.1in}
\Scale[0.78]{
\begin{minipage}{1.0\columnwidth}
\vspace{-0.08in}
\begin{align*}
c'_{W}[1] &\leq c_{W}[1] + \rfmth \;\;\; (\because \text{~Lemma 2}) \\
&\leq \frac{1}{2} \left(\sum_{i=1}^2 c_{W-1}[i] + \rfmth\right) + \rfmth \;\;\; (\because \text{~Lemma 5}) \\
&\leq \frac{1}{3} \left(\sum_{i=1}^{3} c_{W-2}[i] + \rfmth\right) + \sum_{k=1}^{2} \frac{\rfmth}{k} \;\;\; (\because \text{~Lemma 5}) \\
& \hspace{1.5in} \vdots\\
&\leq \frac{1}{n} \left(\sum_{i=1}^{n} c_{W-n+1}[i] + \rfmth\right) + \sum_{k=1}^{n - 1} \frac{\rfmth}{k}\\
&= \frac{1}{n} \left(\sum_{i=1}^{n} c_{W-n+1}[i]\right) + \sum_{k=1}^{n} \frac{\rfmth}{k}\\
\end{align*}
\vspace{-0.3in}
\end{minipage}}

\noindent At this point, we have to apply Lemma 6.\\

\vspace{-0.1in}
\Scale[0.73]{
\begin{minipage}{1.0\columnwidth}
\begin{align*}
c'_{W}[1]
&\leq \frac{1}{N} \left(\sum_{i=1}^{N} c_{W-n}[i] + \rfmth + (N - n) \adth \right) + \sum_{k=1}^{n} \frac{\rfmth}{k}\\
\end{align*}
\end{minipage}}

\noindent Then, we apply Lemma 3 ($k = N$) for $W - n - 1$ times.\\

\Scale[0.68]{
\begin{minipage}{1.0\columnwidth}
\begin{align*}
c'_{W}[1]
&\leq \frac{1}{N} \left(\sum_{i=1}^{N} c_{1}[i] + (W-n) \rfmth + (N - n) \adth \right) + \sum_{k=1}^{n} \frac{\rfmth}{k}\\
\end{align*}
\end{minipage}}

\noindent The maximum value of \begin{footnotesize}$\sum_{i=1}^{N} (c_{j}[i] - c_{j}[N])$\end{footnotesize} is equivalent to that in the proof of Theorem 1. Then, the following holds true.

\Scale[0.62]{
\begin{minipage}{1.0\columnwidth}
\begin{align*} 
c'_{W}[1] - c'_{1}[N]
&\leq \frac{1}{N} \left(\sum_{i=1}^{N} (c_{1}[i] - c_{1}[N]) + (W-n) \rfmth + (N - n) \adth \right)\\
&\;\;\;\;\,+ \sum_{k=1}^{n} \frac{\rfmth}{k}\\
&\leq \frac{1}{N} \left((N-2) \rfmth + (W-n) \rfmth + (N - n) \adth \right)\\
&\;\;\;\;\,+ \sum_{k=1}^{n} \frac{\rfmth}{k}\\
&\leq \frac{1}{N} \left((N-2+W-n) \rfmth + (N - n) \adth \right) + \sum_{k=1}^{n} \frac{\rfmth}{k} \;\;\;  (4)\\
&\hspace{3.8in}
\end{align*}
\end{minipage}}

\noindent Suppose $M'[n]$ is the right side of the inequality (4) for $1 \leq n \leq N$.
Note that $M'[N] = M$, which is the upper bound of $c'_{W}[1] - c_{1}[N]$ when $n$ is greater than $N$ (which is the same as the value when the adaptive refresh is not applied).
Now, we need to find the $n$ value maximizing $M'[n]$.
For $2 \leq n \leq N$, the difference between $M'[n]$ and $M'[n-1]$ is as follows:

\begin{align*}
M'[n] - M'[n-1] = \frac{\rfmth}{n} - \frac{\rfmth + \adth}{N}\\
\end{align*}

\noindent $(M'[n] - M'[n-1])$ is a decreasing function with respect to $n$.
Thus, the largest $n$ (i.e., $n^*$) satisfying $M'[n] >M'[n-1]$ is given by $n^* = \lceil (N \times \rfmth) / (\rfmth+\adth) \rceil$, and it maximizes $M'[n]$.
Finally, we can prove Theorem 2 as follows:

\Scale[0.68]{
\begin{minipage}{1.0\columnwidth}
\begin{align*}
c'_{W}[1] - c'_{W}[N]
&\leq \sum_{k=1}^{n^*} \frac{\rfmth}{k} + \frac{(W-n^*+N-2) \rfmth + (N - n^*) \adth}{N}\\
&= \maxcount'\\
&\!\!\!\!\!\!\!\!\!\!\!\!\!\!\!\!\!\!\!\!\!\!\!\!\!\!\!\!\!\!\!\!\!\!\!\!*\; \maxcount' = \maxcount'[n^*]\\
\end{align*}
\end{minipage}}

\subsection{PARFM Probability of Failure}
\label{sec:parfm_appendix}

A PARA-inspired, intuitive form of a probabilistic prevention scheme, \pararfm, is deployable under the RFM interface.
Whenever an RFM command arrives, \pararfm randomly samples a single aggressor row among the last $\rfmth$ activations and executes the preventive refresh on its victims.
The probability of being selected for a row depends on the ratio of its ACTs on the last $\rfmth$ activations.
\pararfm's protection capability depends on $\rfmth$, which determines the sampling rate.

Failure probability of \pararfm requires two major modifications on the original method of PARA: \emph{the worst-case ACT pattern} and \emph{mathematical formulation}.
First, the number of rows to activate depends on the given $\rfmth$ value, while the worst-case ACT pattern of PARA was to activate a single row continuously.
Suppose only a single row is activated under \pararfm.
In that case, it will always be selected at the next RFM command, and its victims will receive the preventive refresh.
From the attacker's perspective, the cost-effectiveness in minimizing the \pararfm selection and quickly reaching $\nth$ is expressed as the following (where $j$ denotes the number of ACTs for a single row in a single $\rfmth$ activation interval):

\Scale[0.9]{
\begin{minipage}{1.0\columnwidth}
\setcounter{equation}{4}
\begin{align}
\label{eq:cost_effectiveness}
\text{Cost-effectiveness:}\ \left(1-\frac{j}{\rfmth}\right)^{1/j}
\end{align}
\vspace{0.01in}
\end{minipage}}

Because this is a monotonically decreasing function and the $\rfmth$ period that the row is not activated can be ignored (it does not contribute to reaching the $\nth$ ACTs), activating a row only a single time for every $\rfmth$ is the most cost-effective pattern.
We thus base our further formulation on that the $\rfmth$ number of different rows are activated once every $\rfmth$ period.

Compared to PARA, the mathematical formula also must be different.
The following formula is the accurate probability of failure for a single DRAM bank in a tREFW time window.
$Fail(1)$ denotes the failure probability where a single row fails.
$Fail(2)$ denotes the failure probability where two different rows fail in a tREFW window and so on.
Based on Equation~(\ref{eq:cost_effectiveness}), a uniform distribution of ACTs on the activated rows is assumed.

\Scale[0.9]{
\begin{minipage}{1.0\columnwidth}
\begin{align*}
\text{Bank failure probability:}\
\Comb{\Scale[0.65]{\rfmth}}{1}Fail(1)-\Comb{\Scale[0.65]{\rfmth}}{2}Fail(2) \\
+\Comb{\Scale[0.65]{\rfmth}}{3}Fail(3)-\Comb{\Scale[0.65]{\rfmth}}{4}Fail(4)\ \dots \\
\end{align*}
\end{minipage}}

$Fail(1)$ can be calculated using the following recurrence equation where $P[i]$ denotes the failure probability at the i-th RFM command:

\Scale[0.8]{
\begin{minipage}{1.0\columnwidth}
\begin{align*}
P[i] = P[i-1]+\frac{1}{\Scale[0.75]{\rfmth}}(1-\frac{1}{\Scale[0.75]{\rfmth}})^{\nth/2}(1-P[i-\Scale[0.75]{\nth/2}-1]) \\
\end{align*}
\end{minipage}}

The initial condition is as follows.

\Scale[0.8]{
\begin{minipage}{1.0\columnwidth}
\begin{align*}
P[i] = \begin{dcases*}
0 & for $0\ {\leq}\ i\ {\leq}\ {\frac{\nth}{2}}-1$ \\
\left(1-\frac{1}{\rfmth}\right)^{\nth/2} & for $i=\frac{\nth}{2}$ \\
\end{dcases*}
\end{align*}
\vspace{0.01in}
\end{minipage}}

We acquire $Fail(1)$ by calculating the last $P[i]$ in a tREFW window. 
$Fail(2)$ is much smaller than $Fail(1)$ because the $\nth$ value exceeds over 1K even at the most pessimistic RH vulnerability.
The probability of more than one row reaching the $\nth$ ACT value without being refreshed is much less likely compared to that of a single row.
Therefore, we estimate the probability of failure (upper-bound) with only the first term of bank failure probability.
Using the bank failure probability, we can acquire the system failure probability based on the number of banks that can be simultaneously attacked ($N_{banks}$). 

\Scale[0.9]{
\begin{minipage}{1.0\columnwidth}
\begin{align*}
\text{System Failure Probability:}\  1-(1-Fail(1))^{N_{banks}} \\
\end{align*}
\end{minipage}}

In our experimental system of 2 ranks of 32 banks each, a total of 22 banks can be activated satisfying the tFAW constraints.
In Section~\ref{sec:experimental_setup}, we properly set the $\rfmth$ value on each target $\nth$; therefore, our system failure probability is lower than  $10^{-15}$.


\bibliographystyle{IEEEtranS}
\balance
\bibliography{ref}

\end{document}